\RequirePackage{fix-cm}
\documentclass[twocolumn,epjc3]{svjour3}
\smartqed  
\RequirePackage{graphicx}
 \RequirePackage{mathptmx}      
\usepackage{amsmath,amssymb}
\RequirePackage{latexsym}
\RequirePackage[numbers,sort&compress]{natbib}
\RequirePackage[colorlinks,citecolor=blue,urlcolor=blue,linkcolor=blue]{hyperref}
\journalname{Eur. Phys. J. C}
\begin{document}
\sloppy

\title{Influence of target material impurities on physical results in relativistic heavy-ion collisions
}

\titlerunning{Influence of target material impurities on physics results...}        

\author{D. Bana\'{s}\thanksref{addr1}
        \and
        A. Kubala-Kuku\'{s}\thanksref{addr1} 
        \and
        M. Rybczy\'{n}ski\thanksref{e1,addr1} 
        \and
        I. Stabrawa\thanksref{addr1} 
        \and
        G. Stefanek\thanksref{addr1} 
}

\thankstext{e1}{e-mail: maciej.rybczynski@ujk.edu.pl}


\institute{Institute of Physics, Jan Kochanowski University, 25-406 Kielce, Poland \label{addr1}
}

\date{Received: date / Accepted: date}

\newcommand{\agev}{\mbox{$A$~GeV}}               
\newcommand{\gevc}{\mbox{GeV$/c$}}
\newcommand{\gevcc}{\mbox{GeV$/c^2$}}
\newcommand{\mevc}{\mbox{MeV$/c$}}
\newcommand{\mevcc}{\mbox{MeV$/c^2$}}
\newcommand{\gevf}{\mbox{GeV/fm$^3$}}
\newcommand{\fmm}{\mbox{fm$^{-3}$}}
\newcommand{\fmp}{\mbox{1/fm$^3$}}
\newcommand{\fmc}{\mbox{fm$/c$}}
\newcommand{\miuB}{\mbox{$\mu_{B}$}}
\newcommand{\roots}{\mbox{$\sqrt{s_{_{NN}}}$}}
\newcommand{\Bea}{\mbox{$^{7}$Be}}
\newcommand{\Beb}{\mbox{$^{9}$Be}}
\newcommand{\Ar}{\mbox{$^{40}$Ar}}
\newcommand{\Sc}{\mbox{$^{45}$Sc}}
\newcommand{\Xe}{\mbox{$^{129}$Xe}}
\newcommand{\La}{\mbox{$^{139}$La}}
\newcommand{\Pb}{\mbox{$^{208}$Pb}}
\newcommand{\Pba}{\mbox{$^{204}$Pb}}
\newcommand{\Pbb}{\mbox{$^{206}$Pb}}
\newcommand{\Pbc}{\mbox{$^{207}$Pb}}

\maketitle

\begin{abstract}
This paper presents the studies on the influence of the target material impurities on physical observables registered in heavy ion collisions collected by fixed target experiments. It mainly concerns the measures of multiplicity fluctuations which can be used to searches for critical point of strongly interacting matter, e.g. in the {NA61/SHINE} fixed-target experiment at CERN SPS.
The elemental composition of the targets used in the NA61/SHINE experiment was determined applying wavelength dispersive X-ray fluorescence (WDXRF) technique. The influence of measured target impurities on multiplicity distributions and scaled variance was estimated using simulation events. The modification of the standard analysis was proposed to reduce this influence.

\end{abstract}

\section{Introduction}
\label{sec:Intro}
Its well established fact that matter exists in different states. For strongly interacting matter described by Quantum Chromodynamics (QCD) at least three states are expected: normal nuclear matter, hadron gas and a system of deconfined quarks and gluons (Quark Gluon Plasma, QGP). The conjectured QCD phase diagram~\cite{Gupta:2011wh} is usually displayed in the two dimensional diagram - temperature, T versus baryon chemical potential, \miuB. The QCD theory predicts that the phase transition between hadronic phase and quark-gluon plasma at large \miuB\ region is of first order~\cite{deForcrand:2002hgr,Endrodi:2011gv}.  More qualitative results come from lattice QCD calculations which show that in the vicinity to $\mu_{B}=0$ it is a smooth crossover transition between hadronic and QGP phase. Thus a critical point of strongly interacting matter is the end point of the first order phase transition boundary in the phase diagram, at which the transition is of the second order and one cannot distinguish two phases ~\cite{Fodor:2004nz}. Unfortunately, the QCD predictions are to a large extent qualitative, as QCD phenomenology at finite temperature and baryon number is one of the least explored domains of the theory. Especially due to sign problem at finite~\miuB~region, it is difficult to precisely determine the location of the critical point or even its sure existence~\cite{Gavai:2014ela}.

It is very important to explore the QCD phase structure and search for the critical point theoretically and experimentally. From theoretical side, it is very difficult to precisely determine the location of the critical point due to its non-perturbative feature. Many QCD based models have given different results on location of the critical point~\cite{Stephanov:2004wx}, nevertheless most of the models locate it close to the chemical freeze-out line in the SPS energy range. Experimentally, one can investigate the onset of deconfinement and search for the critical point in ion collisions by a scan of a broad region of the QCD phase diagram.  The scan is possible experimentally by varying the energy and the size of colliding nuclei. Such scan with the energy was done by the NA49 experiment using Pb+Pb central collisions. The results from the NA49 experiment suggest that the onset of deconfinement can be indirectly observed in central Pb+Pb interactions at low SPS energy (\roots$\approx$7.6 GeV), where {\roots} is center-of-mass energy per nucleon pair ~\cite{Alt:2007aa}.

It is worth to emphasize, that the experimental search for the critical point of strongly interacting matter is challenging because of the rapid expansion of the hot and dense medium created in ion collisions. To obtain a goal one has to select sensitive observables and signatures of the critical point and one needs to understand non-critical contributions to the experimental observables. In addition, the freeze-out conditions of the matter created in ion-collisions should be close enough to the boundary that the phase transition signals weren't washed out after the expansion. A characteristic property of the second order phase transition at the critical point is the divergence of the susceptibilities. Consequently, important signals of a second-order phase transition at the critical point are large fluctuations, in particular an enhancement of fluctuations of multiplicity of produced particles and their transverse moments~\cite{Stephanov:1999zu} as well as fluctuations of conserved quantities, such as baryon, electric charge and strangeness number. The most efficient way to study the fluctuations of the system created in an ion-collision is to measure an observable on the event-by-event basis and to study the fluctuations over the ensemble of the events. In the analysis of data one needs to apply various techniques to suppress backgrounds and make precise measurements of fluctuations which include the centrality bin width correction, the suppression of volume fluctuations and auto-correlations, the efficiency correction and the estimation of the statistical and systematical uncertainties. In the fixed target experiments like NA61/SHINE there is an additional effect connected with target material impurities. Such impurities have an influence on fluctuations measures by mixing collisions of projectile ion with various nuclei from the target. The effect is strongest in case of measuring collisions of light ions with the target composed of light nuclei with impurities coming from heavy nuclei. The similar effect is expected in opposite case i.e. heavy ion-heavy nucleus collisions with impurities coming from light nuclei. Its obvious that the influence of impurities can be different for particles measured in different rapidity range.

The paper is organized as follows. In section 2 the NA61/SHINE experiment, the detector as well as collected data and targets are shortly described. The methodology of target impurities measurements by Wavelength Dispersive X-ray Fluorescence (WDXRF) and the results are presented in section 3. The simulations done with use of HIJING model are described in section 4 together with the discussion of the impact of target impurities on multiplicity fluctuations measured by the NA61/SHINE experiment. Section 5 contains a brief description of a method allowing to estimate and to reduce the influence of target material impurities on the measured multiplicity distributions. Finally, section 6 contains the summary and conclusions.

\section{The NA61/SHINE experiment}
\label{sec:Exper}
\begin{figure}
 \includegraphics[width=8.5cm,clip=true]{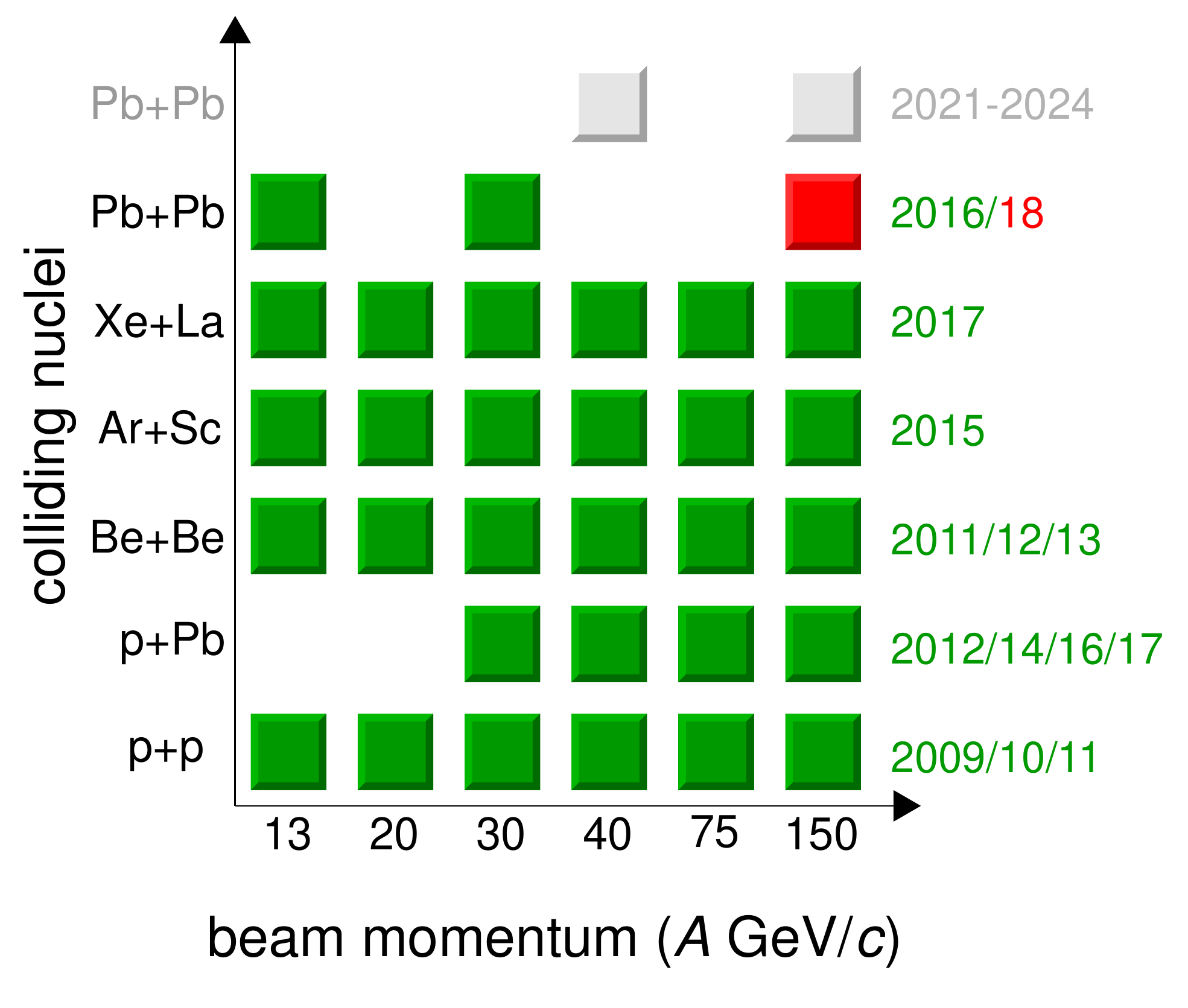}
\caption{The NA61/SHINE data samples for the programme of strong interactions. The recorded data are indicated in green, the approved future data taking in red, whereas the proposed extension for the period beyond 2020 in gray. }
\label{fig:samples}
\end{figure}

NA61/SHINE (SPS Heavy Ion and Neutrino Experiment)~\cite{Antoniou:2006mh,Abgrall:2008zz} is a multi-purpose fix-target experiment to study hadron production in hadron-proton, hadron-nucleus and nucleus-nucleus collisions at the CERN Super Proton Synchrotron (SPS). The strong interaction programme of NA61/SHINE is devoted to the study of the onset of deconfinement and search for the critical point of hadronic matter. The NA49 experiment mainly studied hadron production in Pb+Pb interactions while the NA61/SHINE collects data at varying collision energy and size of the colliding systems. The programme was initiated in 2009 with the p+p colisions. The data samples collected and planned for the future by the NA61/SHINE experiment within the strong interaction program are shown in figure~\ref{fig:samples}.

The NA61/SHINE experiment uses a large acceptance hadron spectrometer located in North Area Hall and the H2 beam-line. The main tracking devices of the spectrometer are large volume Time Projection Chambers (TPCs). Two of them, the vertex TPCs, are located in the magnetic fields of two super-conducting dipole magnets with maximum bending power of 9 Tm. The field in both magnets was lowered in the proportion to the beam momentum in order to optimize the acceptance.  The particles which go through TPCs are almost entirely measured in forward hemisphere (maximal range $-1.0 \lesssim y < y_{beam}$, $y$ is center-of mass rapidity). Two Time of Flight (ToF) counters located on both sides of the beam just behind MTPCs supplement the NA61/SHINE experimental setup. They slightly extend the acceptance region towards midrapidity especially for charged kaons. Other sub-detectors of the NA61/SHINE setup like various beam counters and several other detectors were used in various configurations depending on colliding systems.

During data taking solid targets of \Beb, \Sc, \La, \Pb~and liquid hydrogen target have been used. The solid targets were mounted in the target holder. It has two possible positions for collecting events with target inserted and removed. The latter events were used for the correction of results on off-target interactions. The target parameters are presented in table~\ref{tab:parameters}.

\begin{table}[!ht]
\caption{\label{tab:parameters}
  The parameters of NA61/SHINE solid targets used for the program of strong interactions.}
\begin{center}
\begin{tabular}{l l l l l l l}
\hline
Target   &  Z  &  A  &  Thickness  &  Density  &  Interaction   \\
  &  &  &  (cm)  &  (g/cm$^{3}$)  &  length    \\
  &  &  &  &  &  (g/cm$^{2}$)     \\
\hline
Be     &  4  &    9      &  1.2  &  1.850  &  77.8    \\
Sc     & 21  &   45    &   0.6  &  2.985  &  123.9   \\
La     & 57  &  139   &   0.3, 0.5  &  6.162  &  169.6   \\
Pb     & 82  &  208   &   0.5  &  11.34  &  199.6\\
\hline
\end{tabular}
\end{center}
\end{table}

The experiment measures the event-by-event fluctuations of particle multiplicities, their transverse momenta as well as chemical fluctuations. In this paper we concentrate on the influence of target impurities on multiplicity fluctuations.


\section{Measurement of impurities}
\label{sec:Meas}
\subsection{Sample description}
We analyzed the elemental composition of the beryllium (Be), scandium (Sc) and lanthanum (La) solid samples dedicated to use in the NA61/SHINE experiment. The samples were ordered at professional manufacturer as a high-purity materials (99.99\%). The elemental analysis was performed using samples with a diameter of 40 mm which was prepared by a manufacturer to use it with a commercial WDXRF spectrometer. The samples was delivered in a foil container and did not require additional preparation procedure before the measurement. The lanthanum material was protected from the oxidation process by vacuum packaging. After unpacking the samples were placed directly to the spectrometer and measured.

\subsection{Experimental setup and measurement conditions}
Elemental composition of studied samples were determined applying wavelength dispersive X-ray fluorescence (WDXRF) technique. The physical basis of X-ray fluorescence analysis (XRF) is the photoelectric effect \cite{agarwal,grieken,dziunikowski}. The X-ray primary beam emitted from X-ray tube is directed onto the studied sample. X-ray photons ionize the sample atoms. The excited atoms decay to the ground state emitting characteristic X-ray radiation and photoelectrons (photoelectric effect) and/or electrons as a result of Auger and Coster-Kronig processes. The XRF analytical technique is based on detection, qualitative and quantitative analysis of the characteristic X-rays. Additionally, in WDXRF technique the wavelength dispersive mode of the characteristic X-ray detection is applied. In this detection method the non-monochromatic secondary X-ray beam emitted from the sample is directed onto the crystal on which the X-ray reflection takes place according to Bragg's law. Detector registers monochromatic radiation which wavelength fulfills the Bragg's equation. Finally, the spectrum of the characteristic X-rays is measured, giving the qualitative and quantitative information about the elemental composition of the studied sample \cite{willis}.
In presented studies WDXRF method was used using the AXIOS spectrometer (Panalytical) equipped with an Rh-anode X-ray tube with maximum power of 2.4 kW \cite{panalytical}. The spectrum of the primary excitation X-ray beam is modified by different values of current and voltage of X-ray tube applied in measurements. Additionally, also primary beam filters can be applied: aluminum (200 $\mu$m), aluminum (750 $\mu$m), brass (100 $\mu$m) and brass (400 $\mu$m). The application of the filters results in lowering of radiation background in different energy range improving the detection limit of the WDXRF technique. Use of the brass (400 $\mu$m) filter, due to the reduction of the K series lines of the X-ray tube, allows the determination the rhodium concentration. The attenuation length of the X-ray strongly depends on the its energy and the atomic number of the elements in the sample. For example, for energy 30 keV the attenuation length is 33355  $\mu$m for Be, 785.3 $\mu$m for Sc and 157.5 $\mu$m for La \cite{henke}.
The wavelength dispersive system of the spectrometer uses five crystals (LiF (200), Ge (111), PE (002), PX1 and LiF (220)) which were automatically selected during the measurements. The characteristic X-rays induced in the sample were diffracted on one of the crystals and measured by flow proportional counter for optimal detection of elements up to Fe or a scintillation detector for heavier elements. The measurements were performed in vacuum.
In order to cover the X-ray energy (wavelength) range of the interest it was necessary to perform 11 scans with different current and voltage of the X-ray tube and different diffraction crystal-detector configurations. Detailed information on measurement conditions is presented in the Table \ref{scans}. The parameters of the scans (crystal, detector, primary beam filter, X-ray tube voltage and X-ray tube current) are optimized for the best detection limit of analyzed element. The energy range of characteristic X-ray registered by spectrometer is in the range from 0.5 keV to 36 keV. The lightest element which can be detected with setup configurations possible in AXIOS spectrometer applying element characteristic X-rays is oxygen (O).

\begin{table*}[!ht]
\footnotesize
\caption[1]{Experimental conditions applied in WDXRF measurements: X-ray energy range, K$\alpha$ lines range of elements, L$\alpha$ lines range of elements, crystal, detector, primary beam filter, X-ray tube voltage and X-ray tube current.}
\begin{center}
\label{scans}
\begin{tabular}{c c c c c c c c c }
  \hline
 Number & Energy range & K$\alpha$ lines range & L$\alpha$ lines range & Crystal & Detector & Filter & Voltage & Current \\
 of scan & (keV) & &  &  & & ($\mu$m) & (kV) & (mA)\\
  \hline
  1 & 27-36 & Te-Ce & - & LiF220 & Scint. & brass (100) & 60 & 40\\
  \hline
  2 & 17-29 & Mo-I & - & LiF220 & Scint. & none & 60 & 40\\
  \hline
  3 & 16-29 & Nb-I & - & LiF200 & Scint. & brass (400) & 60 & 40\\
  \hline
  4 & 12-19 & Kr-Tc & Ra-Am & LiF220 & Scint. & Al (750) & 60 & 40\\
  \hline
  5 & 8.5-13.5 & Zn-Rb & Re-U & LiF220 & Scint. & Al (200) & 60 & 40\\
  \hline
  6 & 4.9-8.5 & V-Cu & Pr-W & LiF220 & Flow & none & 50 & 48\\
  \hline
  7 & 3.24-5 & K-V & In-Ce & LiF200 & Flow & none & 24 & 100\\
  \hline
  8 & 1.98-2.66 & P-Cl & Zr-Ru & Ge111 & Flow & none & 24 & 100\\
  \hline
  9 & 1.68-1.80 & Si-Si & Rb-Rb & PE002 & Flow & none & 24 & 100\\
  \hline
  10 & 1.478-1.542 & Al-Al & Br-Br & PE002 & Flow & none & 24 & 100\\
  \hline
  11 & 0.5-1.4 & O-Mg & V-Se & PX1 & Flow & none & 24 & 100\\
  \hline

\end{tabular}
\end{center}
\end{table*}

The quantitative analysis of the spectra was performed with the AXIOS analytical program Omnian \cite{omnian}. In this analysis the uniformity of the sample is assumed. The Omnian package is available for the standardless analysis of all types of samples. Omnian includes advanced algorithms designed to profile known limitations inherent to XRF and includes spectral interference. The $\textit{dark matrix}$ correction provides better accuracy in cases where light elements such as C, H and O contribute to significant absorbance. In generally, corrections which were involved in Omnian quantitative analysis of studied samples were as follows: finite thickness (correction where the sample was not infinite thick for all measured energies) and Compton validation factor (the analysis of unmeasured matrix compounds by using the peak of Compton-scattered primary X-ray beam).
 The certified reference material was always analyzed to validate the analytical procedure before WDXRF measurement of unknown sample. As an example the results of such analysis are presented in the Table \ref{reference_sample} for reference solid sample (Panalytical). In the Table the nominal value of the element concentrations are compared with the experimental values. It can be concluded that, in the range of the experimental uncertainties, the very good agreement was achieved.

\begin{table}[]
\footnotesize
\caption[1]{Comparison of compound concentration in the certified reference solid sample with results obtained using AXIOS spectrometer.}\label{reference_sample}
\begin{center}
\begin{tabular}{c c c }
  \hline
 Compound & nominal & experimental \\
 & concentration (\%) & concentration (\%) \\
  \hline
 Li$_{2}$B$_{4}$O$_{7}$  & 82.7 & 84.8 \\
  \hline
 B$_{2}$O$_{3}$  & 2.5 & - \\
  \hline
 CaO  & 2.80 ($\pm$ 0.02)  & 2.79 ($\pm$ 0.05)\\
  \hline
 Fe$_{2}$O $_{3}$ & 2.00 ($\pm$ 0.02) & 2.02 ($\pm$ 0.04)  \\
  \hline
  P$_{2}$O$_{5}$ & 4.5 ($\pm$ 0.03) & 4.46 ($\pm$ 0.06) \\
  \hline
   SiO$_{2}$ & 4.00 ($\pm$ 0.03) & 4.12 ($\pm$ 0.06)  \\
  \hline
    SrO & 0.50 ($\pm$ 0.01)  & 0.47 ($\pm$ 0.02) \\
  \hline
   ZnO & 1.00 ($\pm$ 0.01) & 0.99 ($\pm$ 0.03) \\
  \hline
\end{tabular}
\end{center}
\end{table}

\begin{figure}[]
\begin{center}
\includegraphics[width=0.4\textwidth]{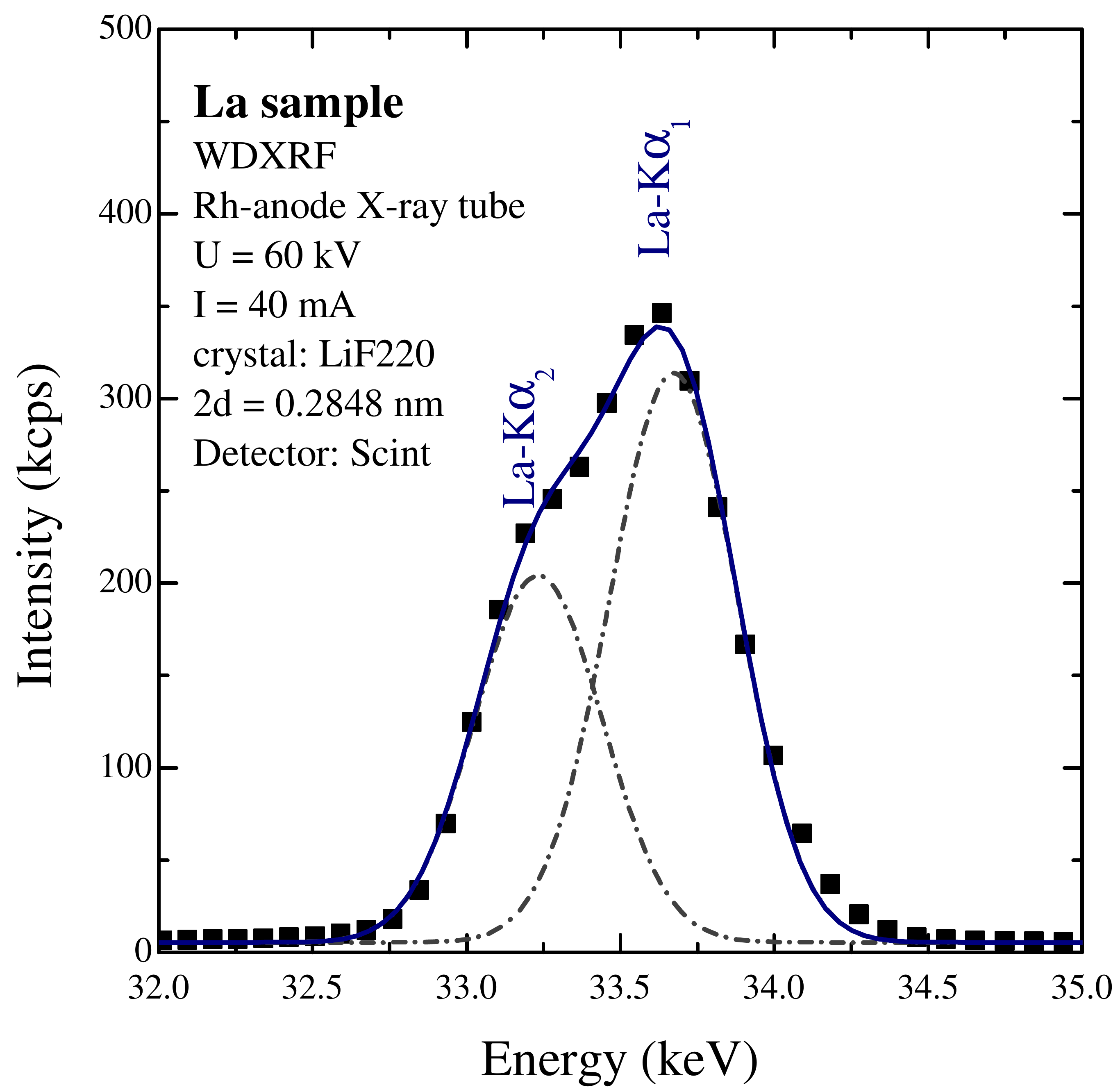}
\end{center}
\caption{Spectrum of the characteristic X-rays emitted from the La sample in the energy range of La-K$\alpha$ lines (La-K$\alpha_{1}$ and La-K$\alpha_{2}$). Spectrum was registered applying WDXRF technique with the experimental conditions given in the Figure.}\label{Figure_La}
\end{figure}

Figure \ref{Figure_La} presents the spectrum of the characteristic X-rays emitted from the La sample in the energy range from 32 keV to 35 keV, corresponding to the La-K$\alpha$ lines (La-K$\alpha_{1}$ and La-K$\alpha_{2}$). The experimental conditions are given inside the figure. The asymmetric shape of the line results from the overlapping of the La-K$\alpha_{1}$ and La-K$\alpha_{2}$ lines. The contribution of the each line was fitted assuming Gaussian profile of the lines and constant full width of the distribution at half of maximum (FWHM). The position of the maximum of the La-K$\alpha_{1}$ line was fitted as 33.673 keV and for La-K$\alpha_{2}$ as 33.233 keV. The width of the each line was 0.490 keV.

\begin{figure}[]
\begin{center}
\includegraphics[width=0.4\textwidth]{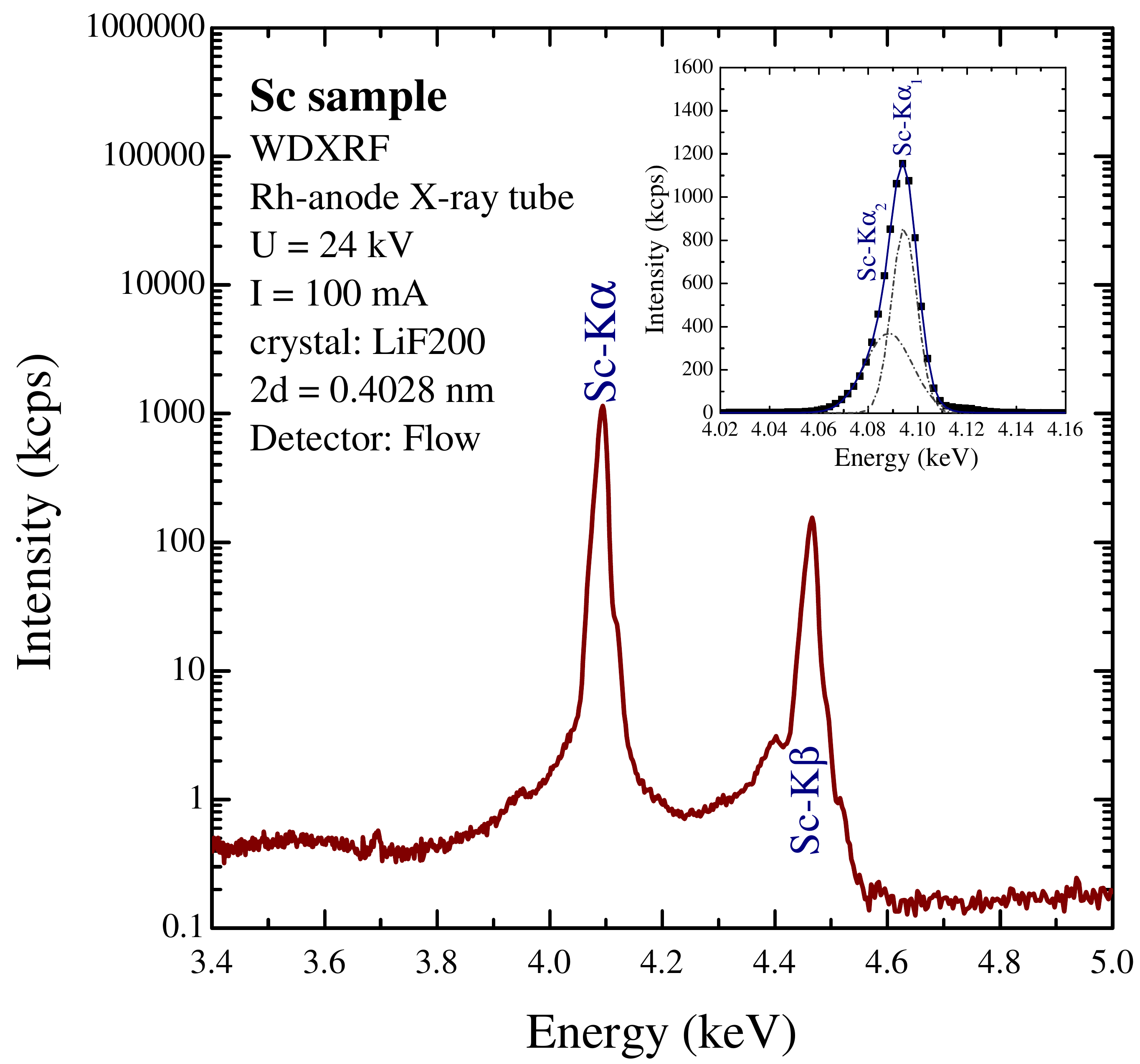}
\end{center}
\caption{Spectrum of the characteristic X-rays emitted from the Sc sample in the energy range of Sc-K$\alpha$ and Sc-K$\beta$ lines (scan 7). Inside the fitted Sc-K$\alpha_{1}$ and Sc-K$\alpha_{2}$ lines are presented. Spectrum was registered applying WDXRF technique with the experimental conditions given in the figure.}\label{Figure_Sc}
\end{figure}

Spectrum presented on the figure \ref{Figure_Sc} was measured for the Sc sample in the range of Sc-K$\alpha$ and Sc-K$\beta$ lines (on presented figure from 3.4 keV to 5 keV). The characteristic X-rays were excited by primary X-ray beam from X-ray tube working with parameters 24 kV and 100 mA and the spectrum was registered using LiF200 crystal and flow detector. Inside the figure the Sc-K$\alpha_{1}$ and Sc-K$\alpha_{2}$ lines are presented. The fitted positions of the maxima of spectral lines are, respectively: 4.095 keV and 4.088 keV, and the FWHM is 0.015 keV, giving the energy resolution on the level 0.37 \%, which is relatively high resolution in spectroscopic application.

The energy resolution of the applied experimental setup was additionally systematically studied for different measurement conditions.
Full widths in the maximum of the intensity (FWHM, $\Delta$E) for characteristic X-ray lines of the elements detected in the studied samples are presented in the Table \ref{FWHM}. For each crystal used in wavelength dispersive mode of X-ray detection the exemplary characteristic line is characterized by its energy in maximum, value of FWHM and energy resolution ($\Delta$E/E). It can be observed that energy resolution changes in very wide range, from 0.13 \% for crystal PE002 and energy 1.489 keV (Al-K$\alpha$) to 1.9 \% for crystal PX1 and energy 0.526 keV (O-K$\alpha$). The high energy resolution allows for unambiguous identification of element intensity even for samples very rich in elemental composition, especially for light elements.

\begin{table}[]
\footnotesize
\caption[1]{Full width in the maximum of the intensity (FWHM, $\Delta$E) for characteristic X-ray lines of the elements detected in the studied samples. The energy (E) of the X-ray lines, the crystal used in wavelength dispersive mode and energy resolution ($\Delta$E/E) are also presented.}\label{FWHM}
\begin{center}
\begin{tabular}{c c c c c c }
  \hline
  Number & Crystal & Line & Energy (E) & FWHM ($\Delta$E) & $\Delta$E/E  \\
  of scan & & & (keV) & (keV) & (\%)\\
  \hline
11 & PX1 & O-K$\alpha$ & 0.526 & 0.010 & 1.90 \\
  \hline
11 & PX1 & F-K$\alpha$ & 0.679 & 0.012 & 1.77 \\
  \hline
10 & PE002  & Al-K$\alpha$ & 1.489 & 0.002 & 0.13 \\
  \hline
9 & PE002  & Si-K$\alpha$ & 1.743 & 0.004 & 0.23 \\
  \hline
8 & Ge111 & S-K$\alpha$ & 2.311 & 0.006 & 0.26 \\
  \hline
7 & LiF200 & Sc-K$\alpha_{2}$ & 4.088 & 0.015 & 0.37 \\
  \hline
7 & LiF200 & Sc-K$\alpha_{1}$ & 4.095 & 0.015 & 0.37 \\
  \hline
6 & LiF220 & Fe-K$\alpha$ & 6.407 & 0.025 & 0.39 \\
  \hline
6 & LiF220 & Ni-K$\alpha$ & 7.481 & 0.036 & 0.48 \\
  \hline
2 & LiF220 & Mo-K$\alpha$ & 17.527 & 0.143 & 0.82 \\
  \hline
1 & LiF220 & La-K$\alpha_{2}$ & 33.233 & 0.490 & 1.47 \\
  \hline
1 & LiF220 & La-K$\alpha_{1}$ & 33.673 & 0.490 & 1.46 \\
  \hline
\end{tabular}
\end{center}
\end{table}

\begin{figure}[]
\begin{center}
\includegraphics[width=0.4\textwidth]{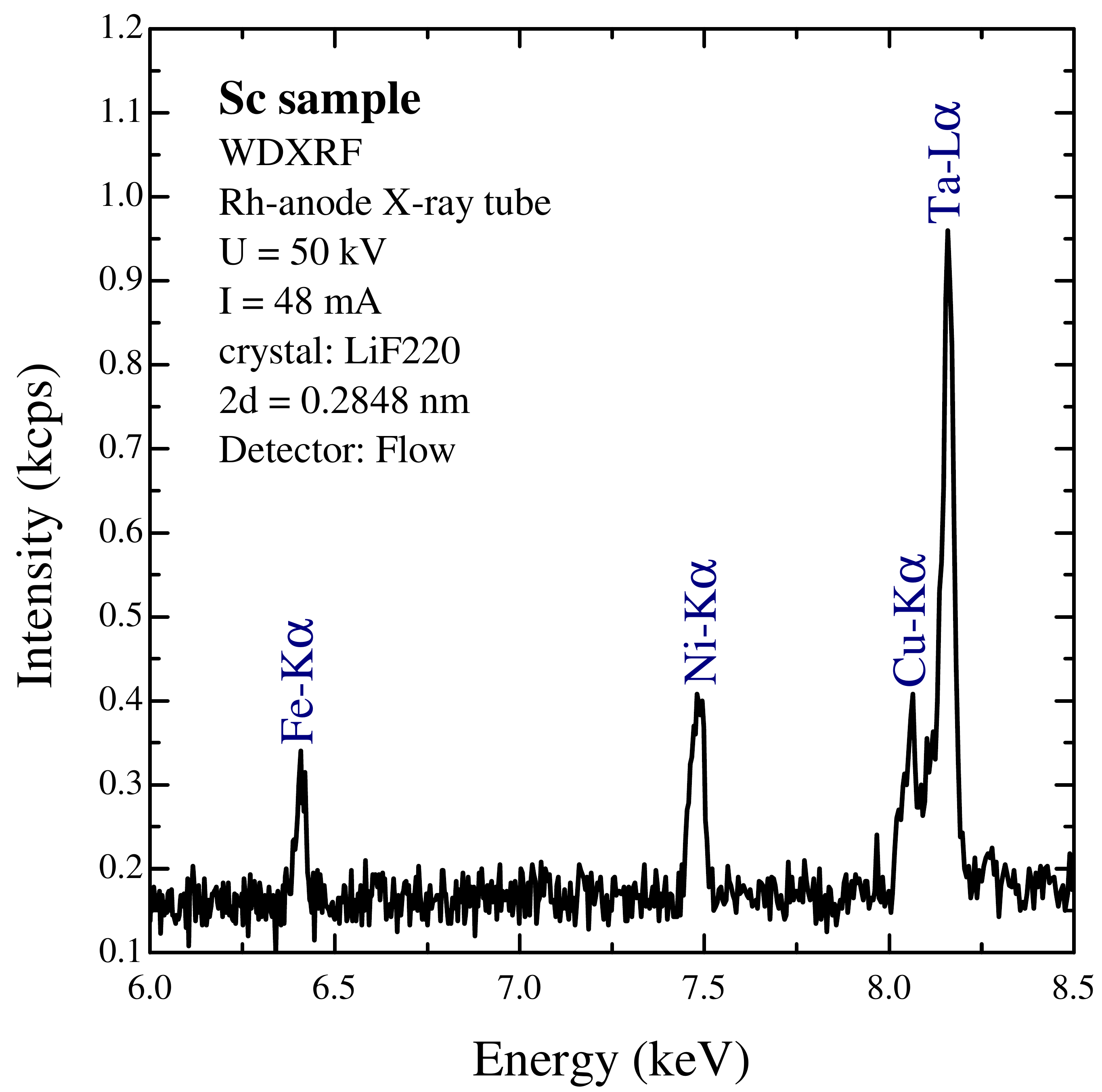}
\end{center}
\caption{Spectrum of the characteristic X-rays emitted from the Sc sample in the energy range from 6 keV to 8.5 keV. On the spectrum the Fe-K$\alpha$, Ni-K$\alpha$, Cu-K$\alpha$ and Ta-L$\alpha$ lines are marked coming from elements being the impurities of the Sc sample. The experimental conditions are also shown in the figure.}\label{Figure_Fe_Sc}
\end{figure}

In presented studies the Be, Sc, La samples were analyzed in the context of the impurity concentrations. For example, figure \ref{Figure_Fe_Sc} presents the spectrum of the characteristic X-rays emitted from the Sc sample in the energy range from 6 keV to 8.5 keV. On the spectrum the Fe-K$\alpha$, Ni-K$\alpha$, Cu-K$\alpha$ and Ta-L$\alpha$ lines are marked coming from elements being the impurities of the Sc sample.

\subsection{Results of the WDXRF measurements}
\label{subsec:Meas_results}

Table \ref{Be_concentration} summarizes the composition of the impurities in the Be sample obtained using WDXRF technique. The following elements were measured: Al, Co, Cu, Fe, Mg, Mn, S, Ti, V, U and W with mass concentration in the range from 0.002 $\%$ (V and W) to 0.13 $\%$ (Fe). The total concentration of admixtured elements in beryllium sample is 0.287 $\%$. The unmeasured Be matrix compound was estimated using the peak of Compton-scattered primary X-ray beam.

In case of the Sc sample the following impurities were detected (table \ref{Sc_concentration}): Al, Bi, Ca, Cl, Cu, Fe, Mg, Ni, Pb, S, Si, Ta, Ti, W and Y. The lowest concentration is 0.002 $\%$ (Y), the highest 0.157 $\%$ (Ti) and total concentration of the impurities is 0.71 $\%$.

Table \ref{La_concentration} summarizes the elemental composition of the La sample. In this sample the main impurity is Zn (0.158 $\%$). The concentration of the rest of the impurities: Al, Ba, Ca, Cl, Fe, Mg, Ni, S, Si, Y, is on the  much lower level, from 0.002 $\%$ to 0.035 $\%$. The total concentration of the admixtured elements is 0.27 $\%$.

The experimental uncertainty of the impurity concentration is calculated from the error of the intensity of the characteristic X-ray and depends on the amount of the element. For the lowest concentration detected in the studied samples the relative uncertainty is on the level 50 $\%$ while for the highest one is about 1 $\%$.

\begin{table}[]
\footnotesize
\caption[1]{Measured elemental composition of Be sample determined using WDXRF technique. Since the lightest element possibly measured by the spectrometer using element characteristic X-rays is oxygen thus Be element was not detected directly. See text for details.}\label{Be_concentration}
\begin{center}
\begin{tabular}{c c }
  \hline
 Element & Mass concentration (\%) \\
   \hline
  Al & 0.039 $\pm$ 0.008 \\
  \hline
  Co & 0.003 $\pm$ 0.002 \\
  \hline
  Cu & 0.008 $\pm$ 0.003 \\
  \hline
  Fe & 0.13 $\pm$ 0.014 \\
  \hline
  Mg & 0.05 $\pm$ 0.008 \\
  \hline
   Mn & 0.027 $\pm$ 0.005 \\
  \hline
  S & 0.003 $\pm$ 0.002 \\
  \hline
  Ti & 0.017 $\pm$ 0.004 \\
  \hline
  V & 0.002 $\pm$ 0.001 \\
  \hline
  U & 0.006 $\pm$ 0.002 \\
  \hline
  W & 0.002 $\pm$ 0.001 \\
  \hline
\end{tabular}
\end{center}
\end{table}

\begin{table}[]
\footnotesize
\caption[1]{Measured elemental composition of Sc sample determined using WDXRF technique.}\label{Sc_concentration}
\begin{center}
\begin{tabular}{c c }
  \hline
 Element & Mass concentration (\%) \\
  \hline
  Al  & 0.144 $\pm$ 0.011 \\
  \hline
  Bi & 0.028 $\pm$ 0.005  \\
  \hline
  Ca & 0.008 $\pm$ 0.003  \\
  \hline
  Cl & 0.011 $\pm$ 0.003 \\
  \hline
  Cu & 0.073 $\pm$ 0.008 \\
  \hline
  Fe & 0.130 $\pm$ 0.011 \\
  \hline
  Mg & 0.008 $\pm$ 0.003 \\
  \hline
  Ni & 0.027 $\pm$ 0.005   \\
  \hline
  Pb & 0.003 $\pm$ 0.002 \\
  \hline
  S & 0.006 $\pm$ 0.002 \\
  \hline
  Sc & 99.29 $\pm$ 1.00 \\
  \hline
  Si & 0.034 $\pm$ 0.006 \\
  \hline
  Ta & 0.065 $\pm$ 0.008 \\
  \hline
  Ti & 0.157 $\pm$ 0.012 \\
  \hline
  W  & 0.017 $\pm$ 0.004 \\
  \hline
  Y & 0.002 $\pm$ 0.001  \\
  \hline
\end{tabular}
\end{center}
\end{table}

\begin{table}[]
\footnotesize
\caption[1]{Measured elemental composition of La sample determined using WDXRF technique.}\label{La_concentration}
\begin{center}
\begin{tabular}{c c }
  \hline
 Element & Mass concentration (\%) \\
  \hline
  Al & 0.018 $\pm$ 0.004 \\
  \hline
  Ba & 0.035 $\pm$ 0.006 \\
  \hline
  Ca & 0.005 $\pm$ 0.002 \\
  \hline
  Cl & 0.006 $\pm$ 0.002 \\
  \hline
  Fe & 0.017 $\pm$ 0.004 \\
  \hline
  La & 99.73 $\pm$ 0.89 \\
  \hline
  Mg & 0.003 $\pm$ 0.002 \\
  \hline
  Ni & 0.002 $\pm$ 0.001 \\
  \hline
  S & 0.006 $\pm$ 0.002 \\
  \hline
  Si & 0.015 $\pm$ 0.004 \\
  \hline
  Y & 0.008 $\pm$ 0.003 \\
  \hline
  Zn & 0.158 $\pm$ 0.012 \\
  \hline
\end{tabular}
\end{center}
\end{table}

The lowest value of given element concentration which can be detected, called low limit of the detection (LLD), using WDXRF spectrometer depends on the one hand on the experimental conditions and on the other hand on the type of the studied sample matrix, and can be calculated using the following formula:
\begin{equation}
\label{eq:1} {LLD=\frac{3\cdot{C}}{I_{n}}\cdot\sqrt{\frac{I_{b}}{t}}},\\
\end{equation}
where C is the element concentration in the studied sample, I$_{n}$ is the net intensity of the characteristic X-ray line, I$_{b}$ is the background level under this line and t is measurement dwell time.
The level of the background is generated mainly by scattering of the X-ray radiation, both the primary beam and secondary characteristic radiation, in the sample. The scattering contribution depends on the energy of the X-rays and on the sample matrix, i.e. element composition in the sample and results in the different values of the element detection limit for as different samples as beryllium (Be), scandium (Sc) and lanthanum (La).

In context of discussed properties of the interaction of the X-ray with matter, the detection limit was estimated using registered X-ray spectra, Tables \ref{Be_concentration}, \ref{Sc_concentration}, \ref{La_concentration} and equation \ref{eq:1}, for all measured samples and for all detected impurities. For Be sample the best achieved detection limit is 3-4 ppm ($\mu$g/g) for U (U-L$\alpha_{1}$, scan number 4) and for S (S-K$\alpha$, scan 8). In case of Sc sample, the lowest value of the LLD was on the level 10 ppm ($\mu$g/g) and it was observed for S element (in general elements detected on the scans 2, 4 and 8). Finally, for sample La, the lowest value of the LLD was on the level 15-20 ppm ($\mu$g/g) and it was observed for S element (scan 8). The presented detection limits of used WDXRF spectrometer is on the sufficient level in the context of presented studies.

\begin{figure}
\includegraphics[width=0.49\textwidth]{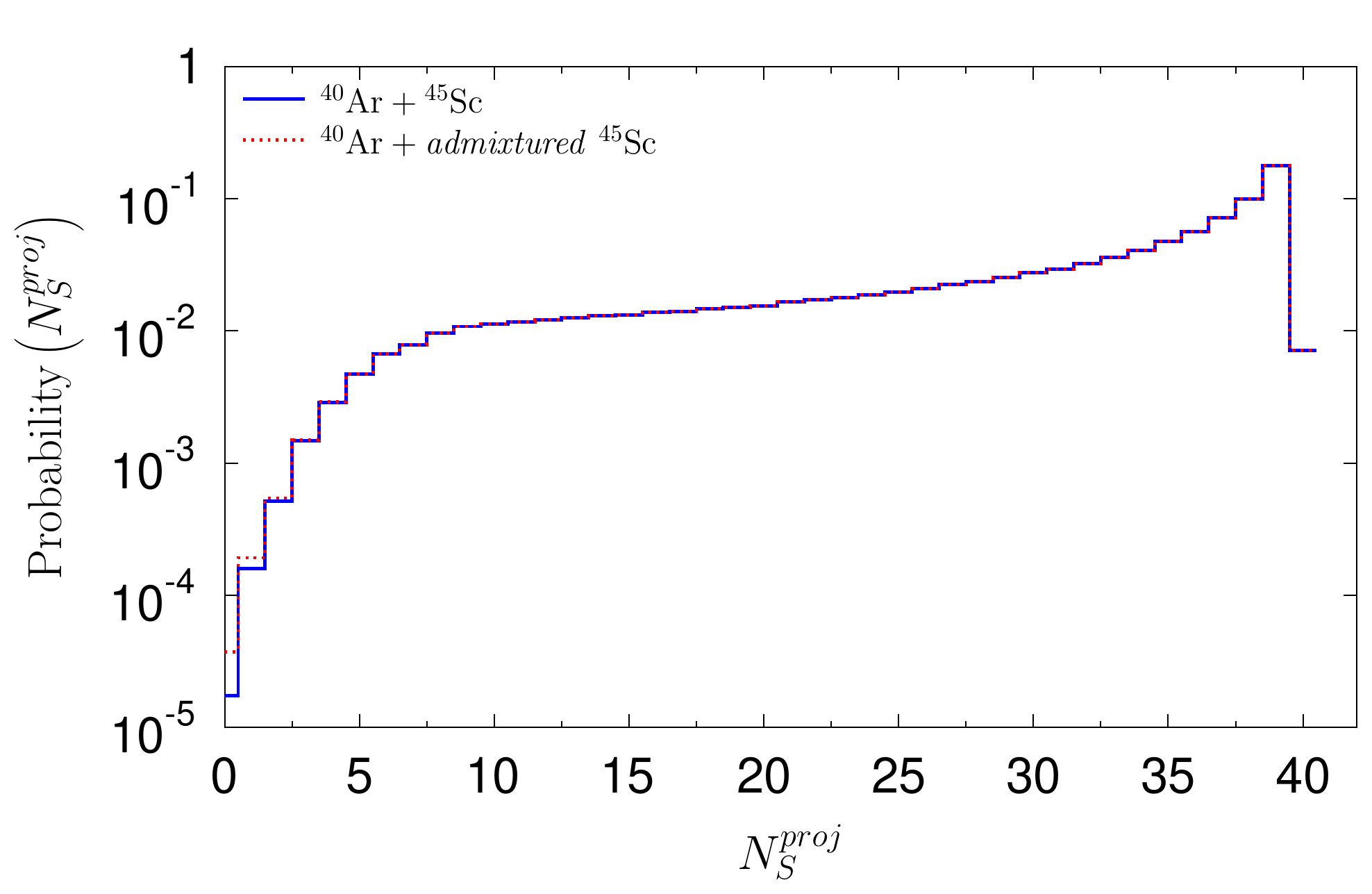}
\caption{(Color online) Distribution of the number of projectile spectators, $N_S^{proj}$ obtained in \Ar+\Sc\ collisions (solid line). Dotted line shows distribution simulated in the collisions of \Ar\ with admixtured Sc target with elements composition taken from table~\ref{Sc_concentration}.}
\label{fig:spec_dist}
\end{figure}

\begin{table}[]
\footnotesize
\caption[1]{Elemental composition of Be target as reported by the target producer.}\label{Be_concentration_prod}
\begin{center}
\begin{tabular}{ c c }
  \hline
 Element & Mass concentration (\%) \\
   \hline
  Be & 99.48  \\
  \hline
  C & 0.15  \\
  \hline
  Mg & 0.08 \\
  \hline
  Al & 0.1 \\
  \hline
  Si & 0.06 \\
  \hline
  Fe & 0.13 \\
  \hline
\end{tabular}
\end{center}
\end{table}

\section{Impact on physical results}
\label{sec:Impact}
This section provides a brief description of our method to estimate an impact of impurities present in target materials on physics observables registered by the NA61/SHINE experiment. We focus on fluctuations of the number of charged particles produced in collisions. The magnitude of these fluctuations if often measured by the scaled variance of multiplicity distribution, $\omega\left(N\right)$, defined as:
\begin{equation}
\omega\left(N\right)=\frac{Var\left(N\right)}{\langle N\rangle},
\end{equation}
where $Var\left(N\right)$ is the variance of the distribution and $\langle N\rangle$ is the average multiplicity.
We also use the relative change of the value of scaled variance, $\Delta$ as a measure of the influence of impurities on multiplicity distributions. It is defined as:
\begin{equation}
\Delta=\frac{|\omega_{pure}\left(N\right)-\omega_{admix}\left(N\right)|}{\omega_{pure}\left(N\right)}\cdot 100\%,
\end{equation}
where $\omega_{pure}\left(N\right)$ is the scaled variance of multiplicity distribution of particles produced in collisions with target containing 100\% of the nominal element ({\it pure target}) and $\omega_{admix}\left(N\right)$ is the one calculated for collisions with target containing impurities ({\it admixtured target}).

\begin{figure*}
\includegraphics[width=0.49\textwidth]{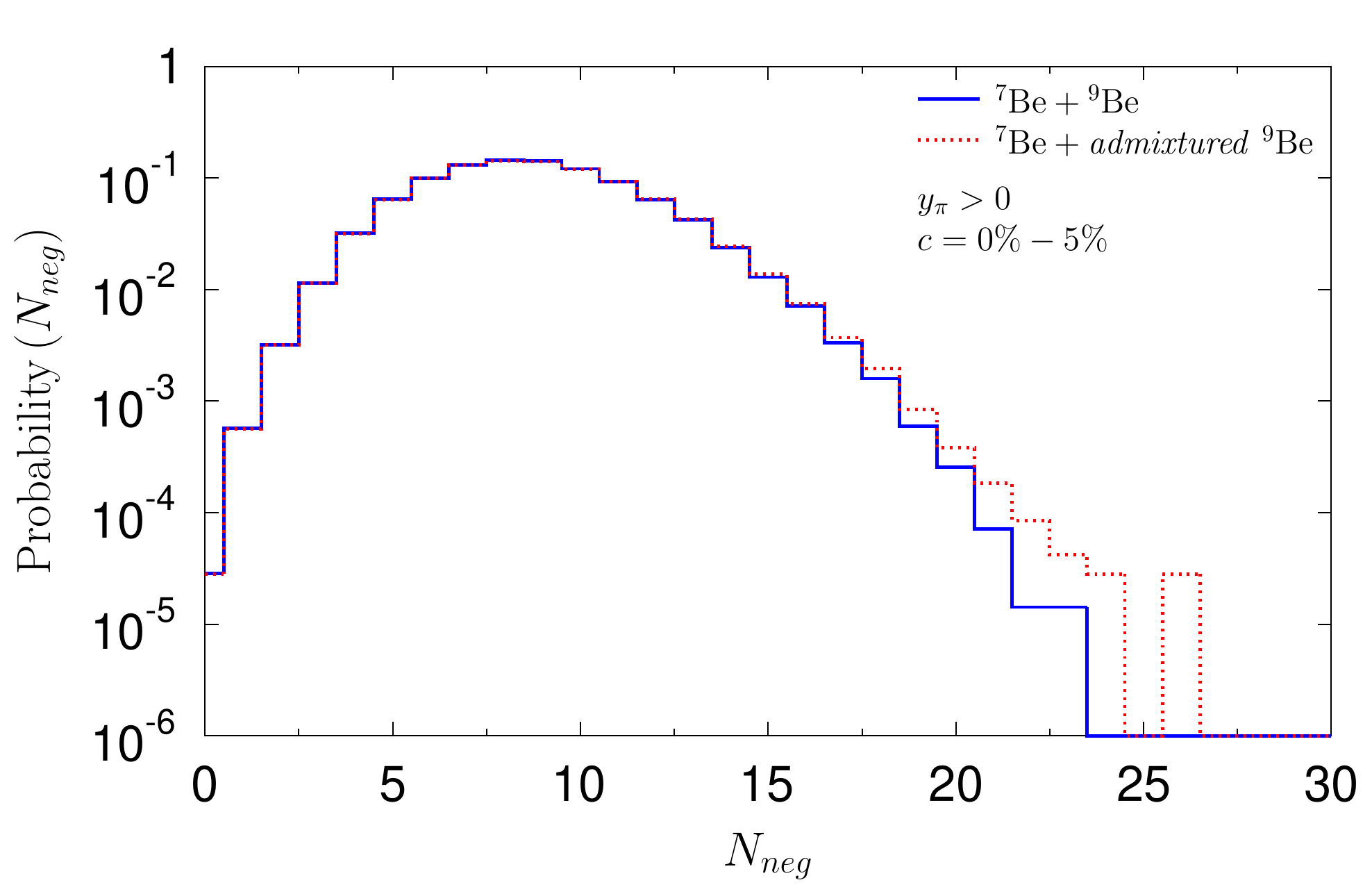}
\includegraphics[width=0.49\textwidth]{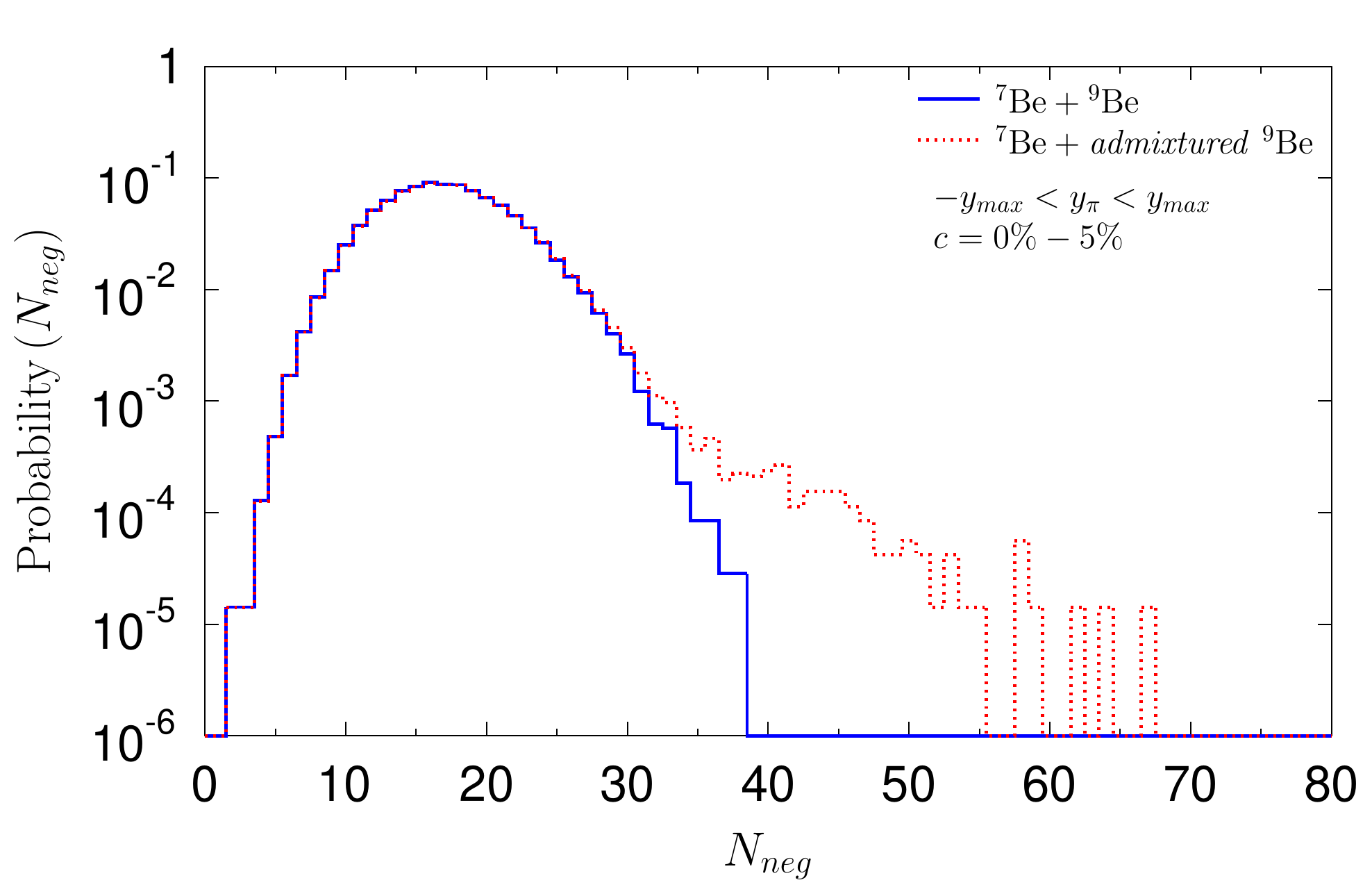}
\caption{(Color online) Multiplicity distributions of negatively charged particles generated in \Bea+\Beb\ collisions in forward (left panel) and full (right panel) kinematical acceptances. Solid lines show distributions generated assuming target composed of 100\% of \Beb\ nuclei whereas with dotted line we present multiplicity distributions resulting from collisions of \Bea\ nucleus with target composed of nuclei according to our WDXRF measurement, table~\ref{Be_concentration}.}
\label{fig:Be_kielce}
\end{figure*}

\begin{figure*}
\includegraphics[width=0.49\textwidth]{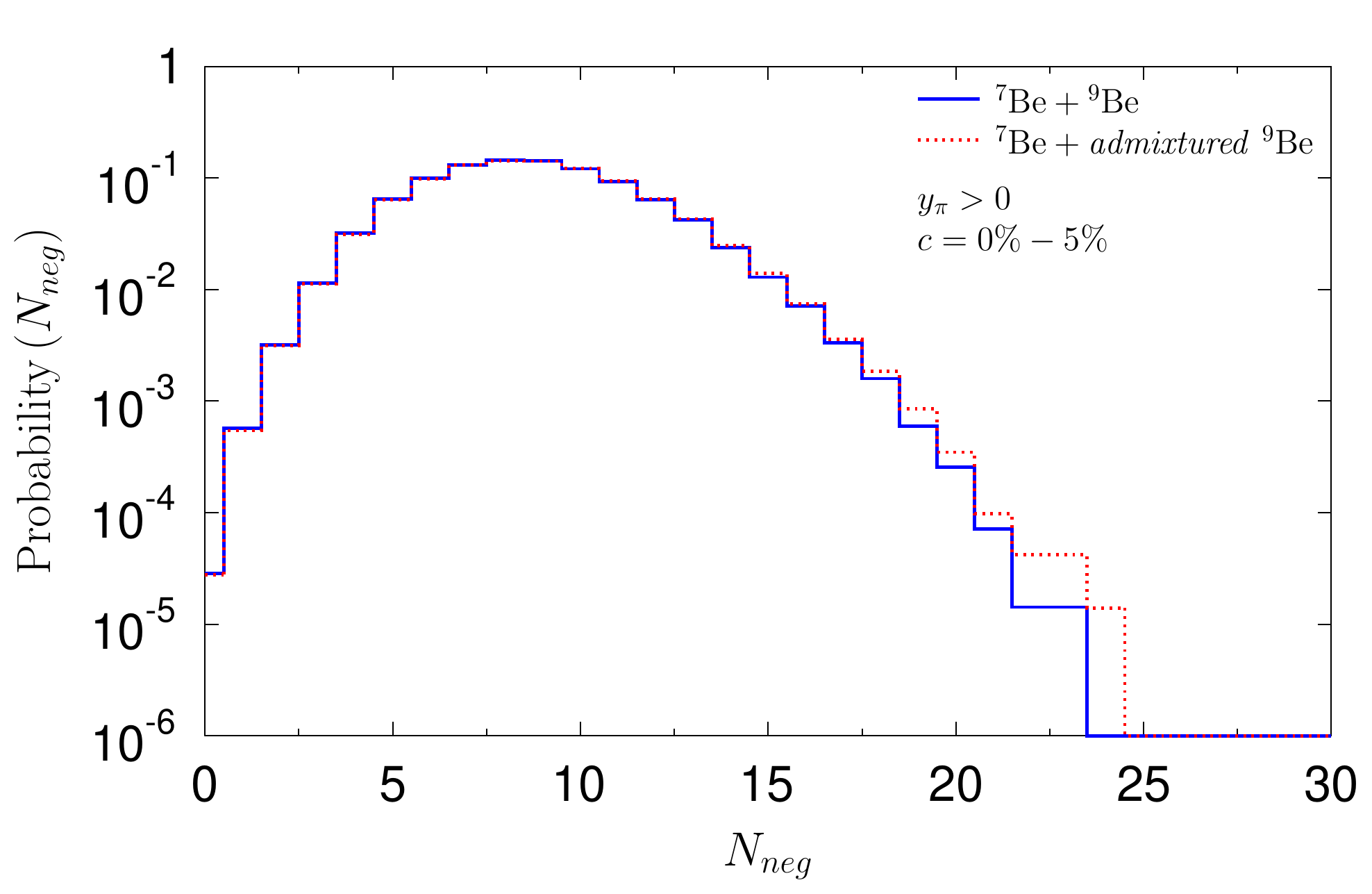}
\includegraphics[width=0.49\textwidth]{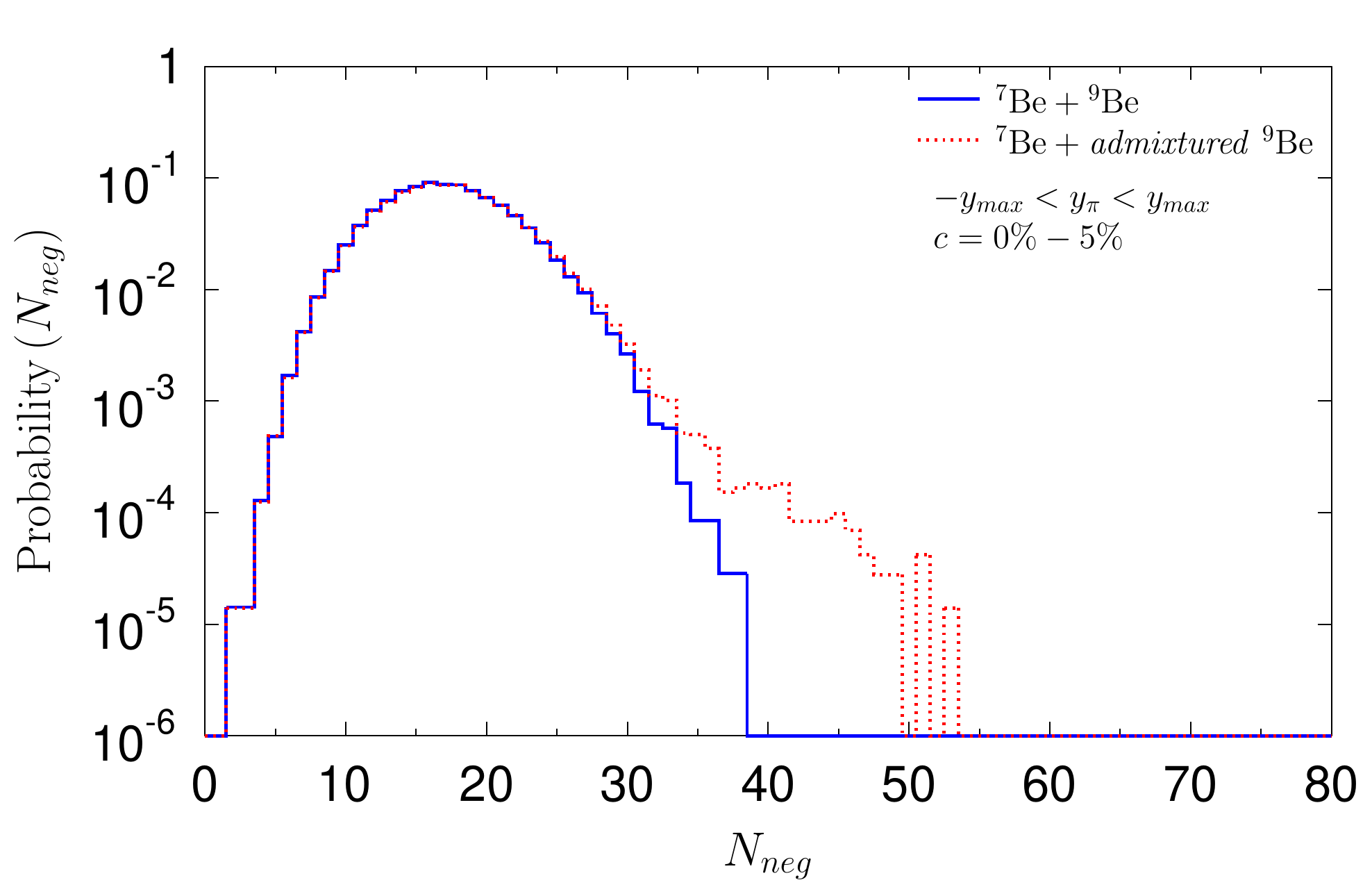}
\caption{(Color online) The same as in figure \ref{fig:Be_kielce} but target was composed of nuclei according to information reported by target producer, see table~\ref{Be_concentration_prod}.}
\label{fig:Be_producer}
\end{figure*}

\subsection{Simulations}
\label{subsec:Sim}

We analyzed the simulated multiplicity distributions of negatively charged particles generated in the collisions of \Bea+\Beb, \Ar+\Sc, \Xe+\La\ and \Pb+\Pb\ at energy measured in laboratory frame, $E_{lab}=150$~GeV/nucleon (\roots$=$17 GeV). Simulated events were generated from HIJING model~\cite{Gyulassy:1994ew}. Two sets of simulation events were build. In both of them the projectile was the same as in the NA61/SHINE experiment, it means \Bea, \Ar, \Xe, and \Pb. In the first set the target entirely consists of nominal elements \Beb, \Sc, \La, \Pb\ ({\it pure target}).  The procedure of preparation of the second set is following. We generated the collisions of given projectile nucleus with all kinds of nuclei present in the {\it admixtured target}, separately. Then we generated set of collisions taking events from simulated collisions with probabilities 
\begin{equation}
p=\frac{N\cdot\sigma_{AB}}{\sum N\cdot\sigma_{AB}},
\end{equation}
where $N$ is the number concentration~\footnote{Number concentrations of target elements was calculated with accordance to the measured mass concentrations of elements described in the previous sections.} of a given element in the admixtured target and $\sigma_{AB}=\pi\cdot R^{2}\left(A^{1/3}+B^{1/3}-\delta\right)^{2}$ with $R=1.4$~fm and $\delta=1.12$ is the collision cross section of projectile nucleus with atomic mass $A$ and target nucleus with atomic mass $B$~\cite{Bradt:1950zza,Westfall:1979zz}. In the case of \Bea+\Beb\ collsions, since the lightest element possibly measured by the AXIOS spectrometer is oxygen thus Be element was not detected and we assumed for simulations the mass concentration of \Beb\ to be equal $100\% -0.287\%=99.713\%$. For \Bea+\Beb\ interactions we also simulated events using information of concentrations of impurities reported by the target producer, see table~\ref{Be_concentration_prod}. To prepare the admixtured Pb target we assumed the number concentrations of nuclei in the target proportional to the abundances of stable Pb isotopes in the Earth's core~\cite{Pb_isotopes}. Namely, we composed the admixtured Pb target of 1.4\% \Pba, 24.1\% of \Pbb, 22.1\% of \Pbc, and 52.4\% of \Pb\ nuclei.

For each simulated event the following quantities were registered: a) the number of projectile spectators; b) the multiplicity of negatively charged particles generated in full kinematical acceptance; c) the multiplicity of negatively charged particles in forward kinematical acceptance, defined by $y_{\pi}>0$. We selected 5\%~\footnote{10\% in the case of \Pb+\Pb\ collisions} most central collisions using number of spectator nucleons from projectile nucleus, $N_S^{proj}$ as a measure of centrality of collision, similarly as used in the NA61/SHINE experiment~\cite{Abgrall:2014xwa}~\footnote{The NA61/SHINE experiment is equipped with the projectile spectator detector, PSD, which is an calorimeter measuring energy carried by the spectator nucleons from projectile, $E_S^{proj}$. The number of spectator nucleons can be estimated as $N_S^{proj}=E_S^{proj}/E_{lab}$, where $E_{lab}$ is the energy carried by single spectator nucleon from projectile. Knowing $N_S^{proj}$ it is straightforward to estimate the number of participating nucleons from projectile, $N_P^{proj}=A-N_S^{proj}$, where $A$ denotes the atomic mass of projectile nucleus.}. Each set of collisions contains $5\cdot 10^{5}$ minimum bias events, thus also the number of central events is enough to limit statistical uncertainties. In figure~\ref{fig:spec_dist} we show the typical distribution of the number of projectile spectators obtained in \Ar+\Sc\ collisions. In figure \ref{fig:spec_dist} we also present $N_S^{proj}$ distribution obtained in the collisions of \Ar~with Sc target admixtured with elements composition taken from table~\ref{Sc_concentration}. We note very small influence of impurities in the admixtured Sc target on the $N_S^{proj}$ distribution (see first bins in figure \ref{fig:spec_dist}) . The cut $N_S^{proj} \leqslant 10$ for the selection of $0\%-5\%$ centrality range is the same for collisions with pure and admixtured targets because $N_S^{proj}$ is the integer number and the difference between distributions in figure \ref{fig:spec_dist} is very small.

\begin{figure*}
\includegraphics[width=0.49\textwidth]{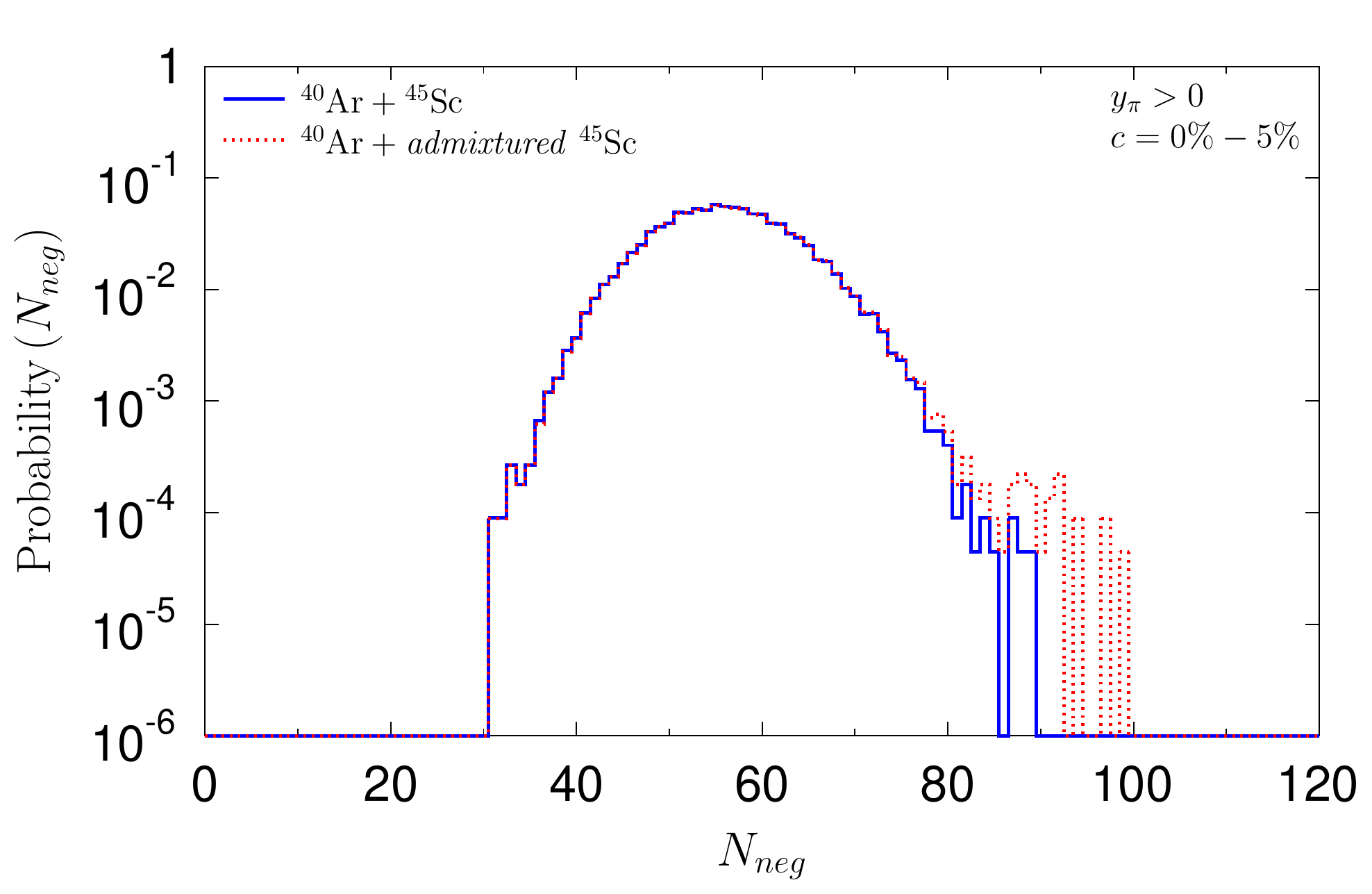}
\includegraphics[width=0.49\textwidth]{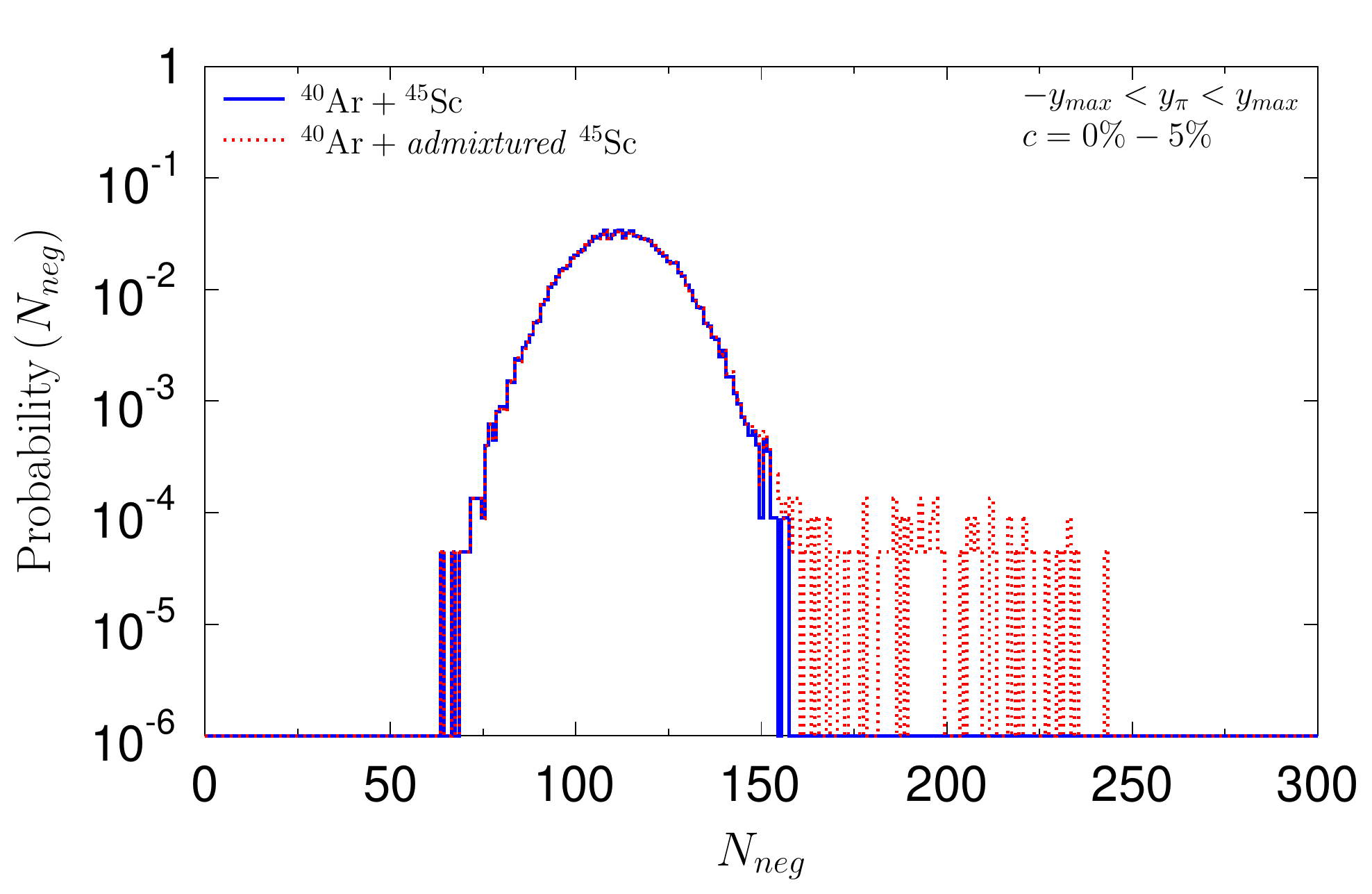}
\caption{(Color online) The same as in figure \ref{fig:Be_kielce} but for \Ar+\Sc\ collisions. Admixtured Sc target was composed of nuclei according to our WDXRF measurement, see table~\ref{Sc_concentration}.}
\label{fig:Sc}
\end{figure*}

\begin{figure*}
\includegraphics[width=0.49\textwidth]{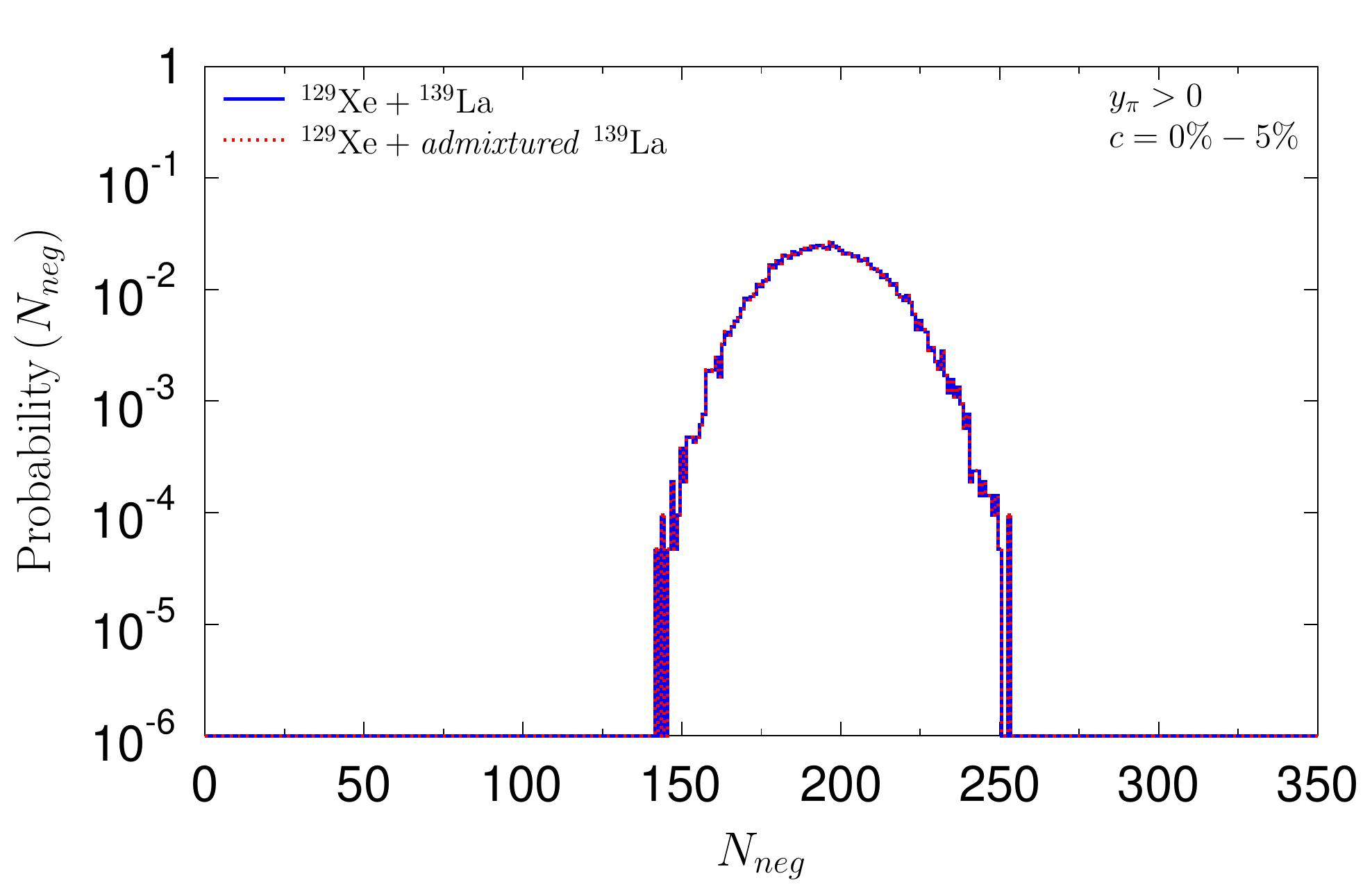}
\includegraphics[width=0.49\textwidth]{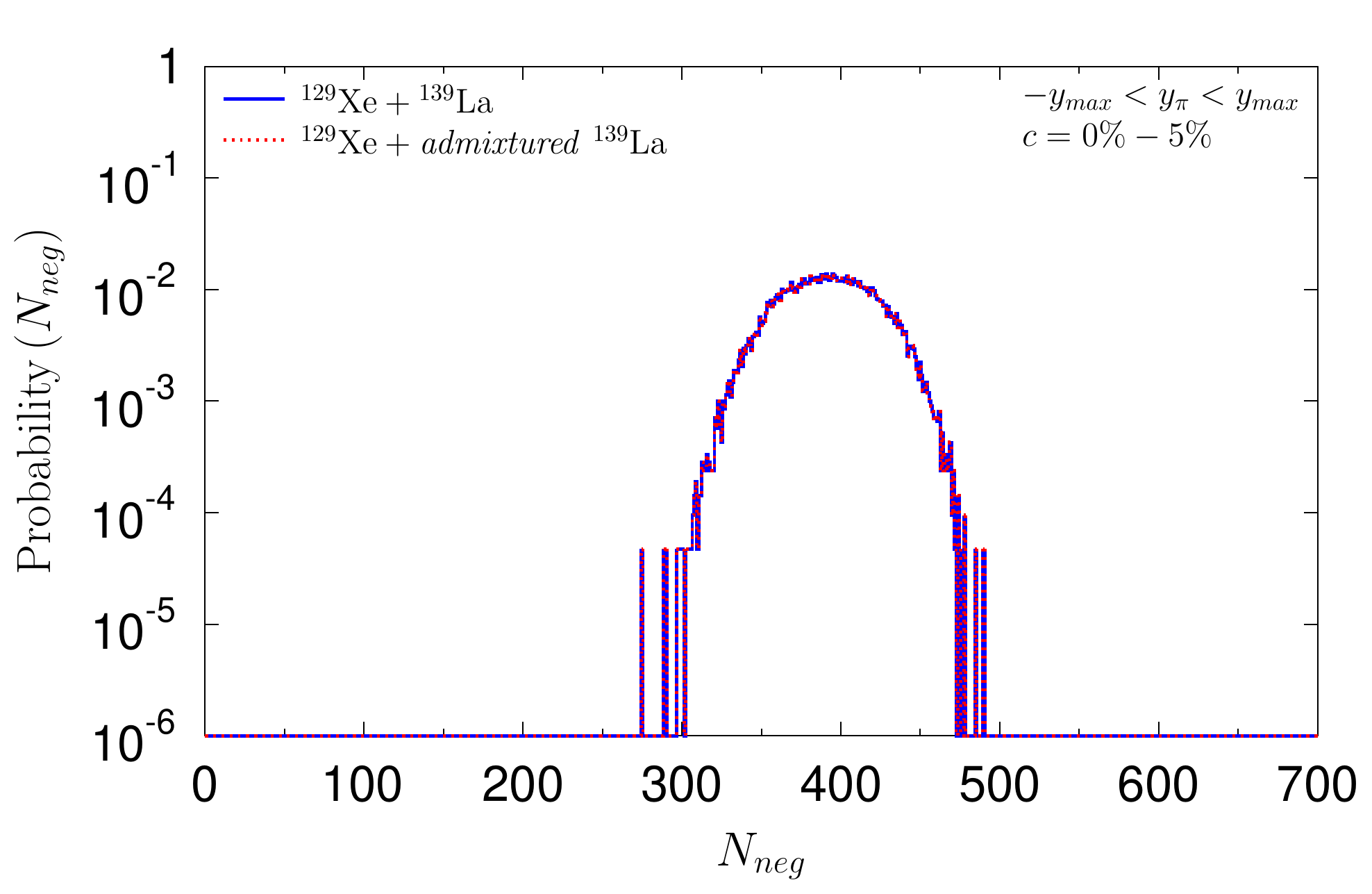}
\caption{(Color online) The same as in figure \ref{fig:Be_kielce} but for \Xe+\La\ collisions. Admixtured La target was composed of nuclei according to our WDXRF measurement, see table~\ref{La_concentration}.}
\label{fig:La}
\end{figure*}

\subsection{Results of the simulations}
\label{subsec:Phys_res}

In this subsection we present the results of the analysis of generated events. We focus on multiplicity distributions of negatively charged particles. Figure~\ref{fig:Be_kielce} contains multiplicity distributions of negatively charged particles generated in \Bea+\Beb\ collisions in forward and full kinematical acceptances. Solid lines show distributions generated assuming target composed of 100\% of \Beb~ nuclei whereas the dotted line represents multiplicity distributions resulting from collisions of \Bea\ nucleus with target composed of nuclei according to our WDXRF measurement, table~\ref{Be_concentration}. In the case of collisions with admixtured target we note a substantial right-hand side tails in both distributions, of particles produced in forward as well as in full kinematical acceptance. The origin of these tails come mainly from the presence of heavy nuclei (V, U, W) in the target. What is also very important the selection of 5\% of most central events favors collisions of \Bea\ with heavier nuclei in target and changes the contribution of different nuclei present in target material to the observed multiplicity distribution. So, in the centrality selected events there is a different contribution of target nuclei than in the target material. The relative change of scaled variance in \Bea+\Beb\ collisions is $\Delta=2.2\%$ and $\Delta=11.5\%$ for forward and full kinematical acceptances, respectively.

Figure~\ref{fig:Be_producer} presents similar results like in figure~\ref{fig:Be_kielce} but obtained for Be target elemental composition reported by the target producer. Similarly, like in figure~\ref{fig:Be_kielce} there are also right-hand side tails in the distributions coming from the collisions with heavy nuclei in admixtured target. Here, the size of the tails is smaller due to the different composition of target elements, in particular lack of very heavy elements, see table~\ref{Be_concentration_prod}. In this case the relative change of scaled variance is equal to $1.4\%$ and $8.7\%$ for forward and full kinematical acceptances, respectively.

\begin{figure*}
\includegraphics[width=0.49\textwidth]{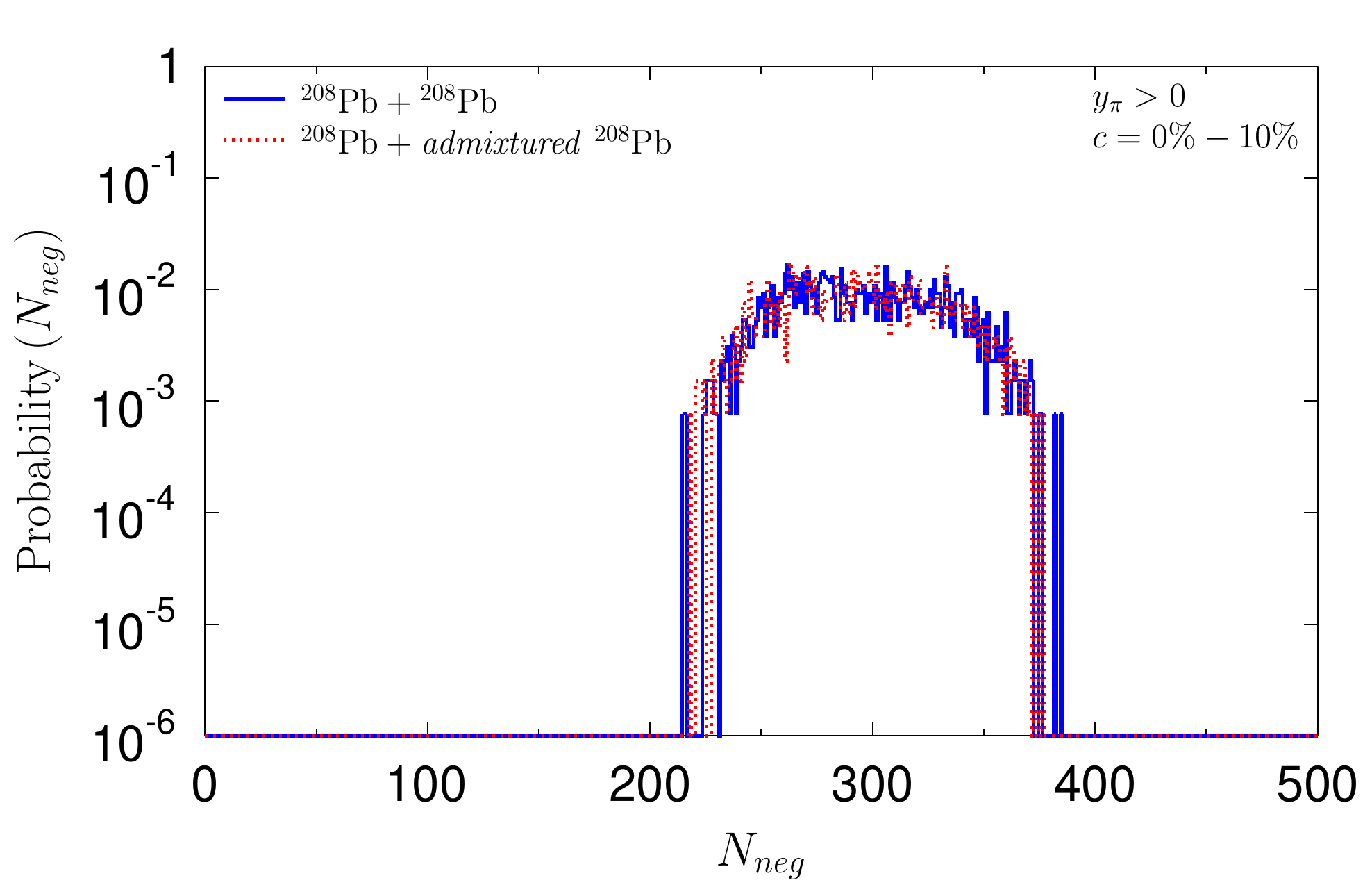}
\includegraphics[width=0.49\textwidth]{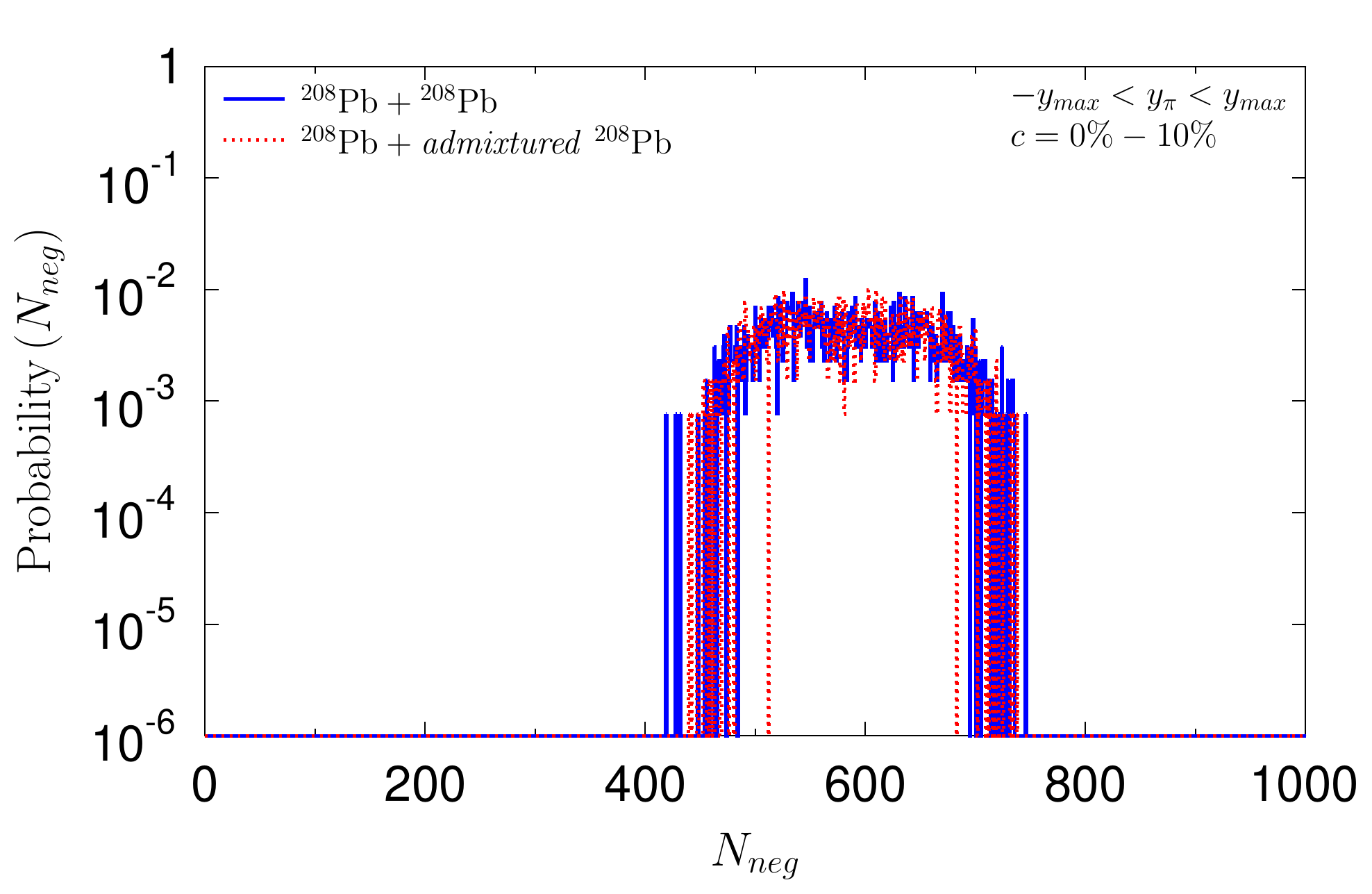}
\caption{(Color online) Multiplicity distributions of negatively charged particles, generated in \Pb+\Pb\ collisions in forward (left panel) and full (right panel) kinematical acceptances. Full lines show distributions generated assuming target composed of 100\% of \Pb\ nuclei whereas using dotted line we present multiplicity distributions coming from collisions with target composed of nuclei proportional to the abundances of stable Pb isotopes in the Earth's core.}
\label{fig:Pb}
\end{figure*}

Figures~\ref{fig:Sc}-\ref{fig:La} contain multiplicity distributions of negatively charged particles generated in the collisions of \Ar+\Sc\ and \Xe+\La.~For \Ar+\Sc~collisions we also report long tails in the distributions and the influence of target impurities are even larger than in the case of \Bea+\Beb\ collisions. The relative change of scaled variance in \Ar+\Sc\ collisions is $\Delta=3.9\%$ and $\Delta=16.4\%$ for forward and full kinematical acceptances, respectively. The presence of Tantalum and Tungsten nuclei in the Sc target is mostly responsible for the asymmetric widening of multiplicity distributions. In contrast, for \Xe+\La\ collisions there is no influence of impurities for the analyzed multiplicity distributions. As one can deduce from figure \ref{fig:La} the relative change of the value of scaled variance in \Xe+\La\ collisions is $\Delta\approx 0.0\%$ for both forward and full kinematical acceptances. Note that the presence of impurities with atomic mass lower than dominating element in the sample does not affect multiplicity distributions mainly due to the selection of 5\% of most central events.

Figure~\ref{fig:Pb} shows multiplicity distributions of particles generated in 10\% most central events of \Pb+\Pb\ collisions. Here for the simulation of admixtured target we assumed the presence of stable Pb isotopes proportional to their abundances in the Earth's core~\cite{Pb_isotopes}, as described in in subsection~\ref{subsec:Sim}. We do not see any substantial influence of Pb isotopes for discussed multiplicity distributions. The relative change of the value of scaled variance in Pb+Pb collisions is, $\Delta\approx0.0\%$ and
$\Delta=2.8\%$ for forward and full kinematical acceptances, respectively. We also performed similar simulations for lower energy, $E_{lab}=40$~GeV/nucleon ($\sqrt{s_{NN}}=8.8$~GeV) with very similar results as for the collisions at $E_{lab}=150$~GeV/nucleon.

The results are summarized in tables~\ref{summ_forward_acc}-\ref{summ_full_acc} where we present numerical values of average multiplicity and corresponding scaled variance of presented multiplicity distributions in forward and full kinematical acceptances, respectively.

\begin{table}[]
\footnotesize
\caption[1]{Average multiplicity of negatively charged particles and corresponding scaled variance of multiplicity distribution of particles generated in forward kinematical acceptance.}\label{summ_forward_acc}
\begin{center}
\begin{tabular}{c c c}
  \hline
 Colliding system & $\langle N_{neg}\rangle$ & $\omega\left(N_{neg}\right)$ \\
   \hline
  \Bea+\Beb & 8.74$\pm$ 0.01 & 0.864$\pm$ 0.005 \\
  \hline
  \Bea+{\it admix.} \Beb [Producer] & 8.79$\pm$ 0.01 & 0.876$\pm$ 0.005  \\
  \hline
  \Bea+{\it admix.} \Beb [WDXRF] & 8.78$\pm$ 0.01 & 0.883$\pm$ 0.005\\
  \hline
  \Ar+\Sc & 56.02$\pm$ 0.05 & 0.92$\pm$ 0.01 \\
  \hline
  \Ar+{\it admix.} \Sc & 56.11$\pm$ 0.05 & 0.956$\pm$ 0.009 \\
  \hline
  \Xe+\La & 195.5$\pm$ 0.1 & 1.3$\pm$ 0.012 \\
  \hline
  \Xe+{\it admix.} \La & 195.5$\pm$ 0.1 & 1.3$\pm$ 0.012 \\
  \hline
  \Pb+\Pb  & 296.0$\pm$ 0.2 & 3.8$\pm$ 0.05\\
  \hline
  \Pb+{\it admix.} \Pb & 295.4$\pm$ 0.2 & 3.8$\pm$ 0.05\\
  \hline
\end{tabular}
\end{center}
\end{table}

\begin{table}[]
\footnotesize
\caption[1]{Average multiplicity of negatively charged particles and corresponding scaled variance of multiplicity distribution of particles generated in full kinematical acceptance.}\label{summ_full_acc}
\begin{center}
\begin{tabular}{c c c}
  \hline
 Colliding system & $\langle N_{neg}\rangle$ & $\omega\left(N_{neg}\right)$ \\
   \hline
  \Bea+\Beb & 17.18$\pm$ 0.02 & 1.139$\pm$ 0.006 \\
  \hline
  \Bea+{\it admix.} \Beb [Producer] & 17.35$\pm$ 0.02 & 1.238$\pm$ 0.006  \\
  \hline
  \Bea+{\it admix.} \Beb [WDXRF] & 17.33$\pm$ 0.02 & 1.27$\pm$ 0.006\\
  \hline
  \Ar+\Sc & 112.1$\pm$ 0.08 & 1.295$\pm$ 0.012 \\
  \hline
  \Ar+{\it admix.} \Sc & 112.45$\pm$ 0.09 & 1.507$\pm$ 0.014 \\
  \hline
  \Xe+\La & 392.1$\pm$ 0.2 & 2.08$\pm$ 0.02 \\
  \hline
  \Xe+{\it admix.} \La & 392.1$\pm$ 0.2 & 2.08$\pm$ 0.02 \\
  \hline
  \Pb+\Pb & 585.6$\pm$ 0.4 & 7.3$\pm$ 0.04 \\
  \hline
  \Pb+{\it admix.} \Pb & 584.0$\pm$ 0.4 & 7.1$\pm$ 0.04 \\
  \hline
 \end{tabular}
\end{center}
\end{table}

\begin{figure*}
\includegraphics[width=0.49\textwidth]{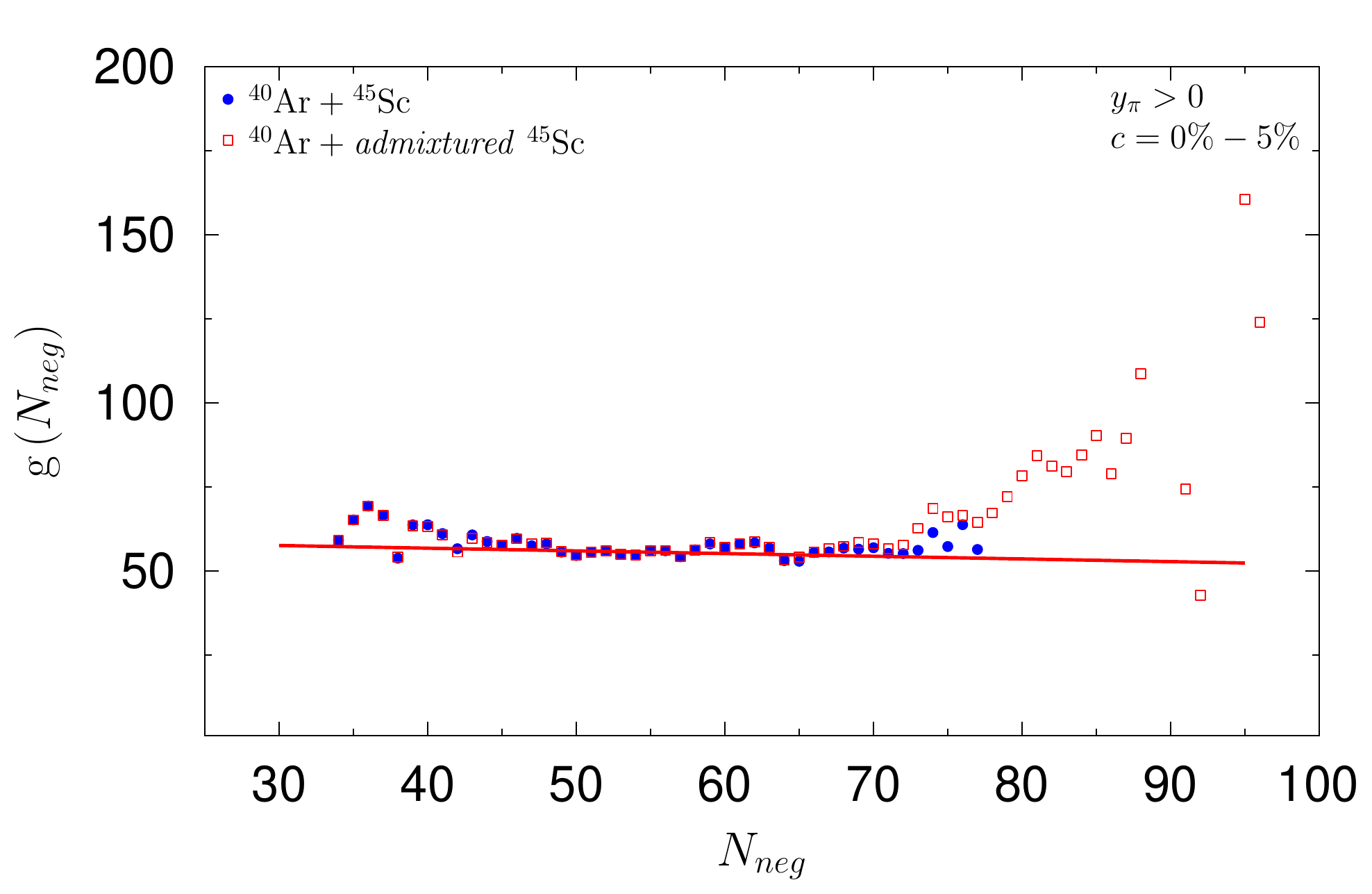}
\includegraphics[width=0.49\textwidth]{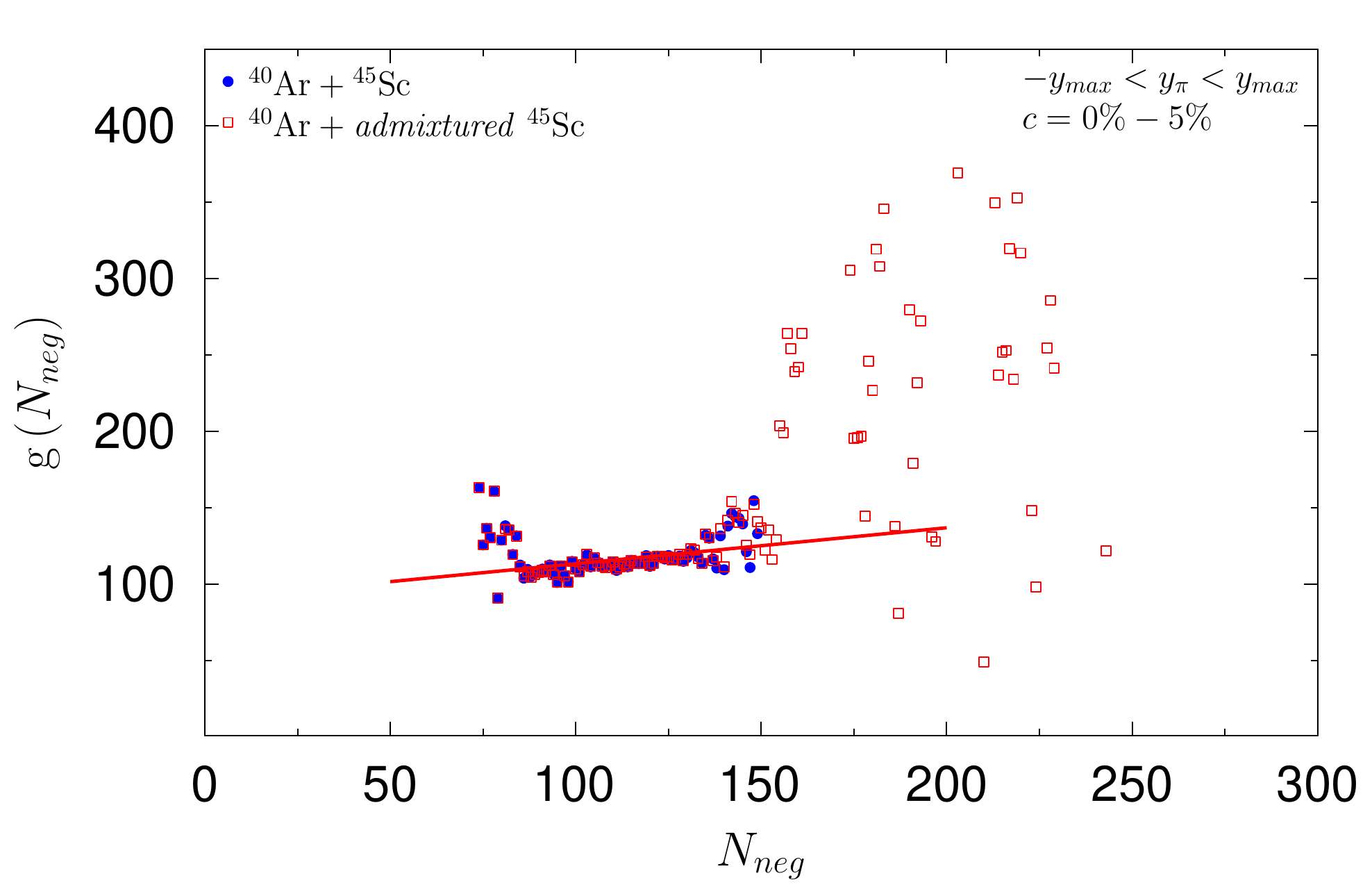}
\caption{(Color online) Recurrence functions $g\left(N_{neg}\right)$ calculated from multiplicity distributions of negatively charged particles generated in \Ar+\Sc\ collisions in forward (left panel) and full (right panel) kinematical acceptances. Full circles show results for collisions with target composed of 100\% of \Sc\ nuclei whereas with empty squares we present results from collisions of \Ar\ nucleus with target composed of nuclei according to our WDXRF measurement, table~\ref{Sc_concentration}. The full lines correspond to the linear functions: $g\left(N_{neg}\right)=60-0.08\cdot N_{neg}$ and $g\left(N_{neg}\right)=90+0.235\cdot N_{neg}$ for forward and full kinematical acceptances, respectively.}
\label{fig:Sc_sol}
\end{figure*}

\section{The method of target impurities influence reduction }
\label{sec:Sol}

This section describes the analysis method which can be used to estimate and reduce the influence  of unwanted collisions caused by the target material impurities.

Let $P\left(N\right)$ be the probability distribution function of multiplicity $N$. Using $P\left(N\right)$ one can define a function $g\left(N\right)$ which satisfies recurrence relation:
\begin{equation}
g\left(N\right)=\frac{\left(N+1\right)P\left(N+1\right)}{P\left(N\right)}.
\end{equation}
Different functions $g\left(N\right)$ describe different multiplicity distributions. For $g\left(N\right)=a={\rm const}$ the corresponding $P\left(N\right)$ equals
\begin{equation}
P\left(N\right)=\frac{a^{N}}{N!}\exp\left(-a\right),
\end{equation}
which is Poisson distribution with the average value $\langle N\rangle=a$. If $g\left(N\right)=a+b\cdot N$ then
\begin{align}
P\left(N\right)&=\frac{k\left(k+1\right)...\left(k+N-1\right)}{N!}\times \nonumber \\ & \times \left(\frac{\langle N\rangle/k}{1+\langle N\rangle/k}\right)^{N}\frac{1}{\left(1+\langle N\rangle/k\right)^{k}}
\end{align}
is a well known negative binomial distribution with $\langle N \rangle$ being average multiplicity and parameter $k=a/b$. In the simplest case: $a=b$ (what leads to $k=1$) one gets 
\begin{equation}
g\left(N\right)=\frac{\langle N\rangle}{\langle N\rangle+1} \left(N+1\right),
\end{equation}
which is commonly known Bose-Einstein enhancement or stimulated emission. Note, that for the case of negative binomial distribution scaled variance of the multiplicity distribution may be expressed as:
\begin{equation}
\omega\left(N\right)-1=\frac{\langle N\rangle}{k}=\frac{b\langle N\rangle}{a}.
\end{equation}
As an example, in figure~\ref{fig:Sc_sol} we present the results obtained for \Ar+\Sc\ collisions. The linear functions $g\left(N_{neg}\right)=a+b\cdot N_{neg}$ with the coefficients $a=60$, $b=-0.08$ and $a=90$, $b=0.235$ for forward and full kinematical acceptances, respectively, describe well the results coming from the {\it pure} \Ar+\Sc\ collisions. Red points scattered randomly on the right-hand sides of the plots correspond to the long tails of the multiplicity distributions resulting from \Ar\ collisions with heavier impurities in the Sc target. In order to avoid the influence of the long, unwanted tails of the multiplicity distributions one can cut them starting from multiplicities where the random scattering of the values of recurrence function $g\left(N\right)$ occurs.

\section{Conclusions}
\label{sec:Conc}
Distributions of charged particles are observables closely connected with the search of critical point of strongly interacting matter already performed in many existing high-energy physics experiments. Part of them, like NA61/SHINE, use fixed-target and dedicated detectors. We analyzed the influence of target material impurities on multiplicity distributions of charged particles produced in most central relativistic heavy-ion collisions using HIJING event generator. The following systems were studied: \Bea+\Beb, \Ar+\Sc, \Xe+\La\ and \Pb+\Pb\ at energies available at CERN SPS. The collisions of considered systems was already registered by the NA61/SHINE experiment at CERN SPS energies. The element compositions of the NA61/SHINE target samples were determined applying wavelength dispersive X-ray fluorescence (WDXRF) technique. Our main results are as follows:

\begin{itemize}
\item In the case of 5\% most central \Bea+\Beb\ interactions we found a substantial influence of target impurities on multiplicity distributions of negatively charged particles produced in both forward and full kinematical acceptances. The presence of long tails in the multiplicity distributions is caused by the contamination of target material with heavy nuclei including Uranium. Selection of central collisions favors heavier nuclei in target and changes the contribution of different nuclei present in target material to the observed multiplicity distributions.

\item In 5\% most central \Ar+\Sc\ collisions we found a large impact of target admixtures on analyzed multiplicity distributions. The relative change of scaled variance is $\Delta=3.9\%$ and $\Delta=16.4\%$ for forward and full kinematical acceptances, respectively. We identified Tantalum and Tungsten nuclei present abundantly in the \Sc\ target which are mostly responsible for the widening of multiplicity distributions.

\item However, in contrast, in the case of central \Xe+\La\ collisions there is no influence of measured target impurities on observed multiplicity distributions.  In La target there is no measured elements heavier than \La\ and the selection of 5\% of most central events practically excludes the contribution of particles produced in collisions of \Xe\ with lighter elements to observed multiplicity distributions.

\item The influence of stable Pb isotopes possibly present in the Pb target on the observed multiplicity distributions was checked. Multiplicity distributions of particles produced in \Pb+\Pb\ collisions were compared with those produced in collisions of \Pb\ nuclei with the target for which we assumed the presence of isotopes  proportional to their abundances in the Earth's core. We report no substantial differences between analyzed multiplicity distributions prepared from different sets of collisions.

\item We proposed the method to estimate and reduce the influence of target impurities on multiplicity analysis and scaled variance calculation.

\end{itemize}

\section*{Acknowledgements}
This work was motivated by the CERN NA61/SHINE experiment programme of study of phase diagram of strongly interacting matter. We thank Antoni Aduszkiewicz, Marek Ga\'{z}dzicki and Zbigniew W\l odarczyk for useful discussions. The equipment was purchased thanks to the financial support of the European Regional Development Fund in the framework of the Polish Innovative Economy Operational Program (contract no. WNP-POIG.02.02.00-26-023/08) and the Development of Eastern Poland Program (contract no. POPW.01.01.00-26-013/09-04). The numerical simulations were carried out in laboratories created under the project ``Development of research base of specialized laboratories of public universities in Swietokrzyskie region'', POIG 02.2.00-26-023/08, 19 May 2009.\\
MR was supported by the Polish National Science Centre (NCN) grant 2016/23/B/ST2/00692.


\end{document}